%% file: paper.tex
\documentclass[10pt,journal,compsoc]{IEEEtran}

\input{meta/packages}
\input{meta/macros}

\begin{document}

\setlength{\pdfpageheight}{\paperheight}
\setlength{\pdfpagewidth}{\paperwidth}

\def\sectionautorefname{Section}
\def\subsectionautorefname{Section}
\def\subsubsectionautorefname{Section}

\title{A Systematic Evaluation of\\Static API-Misuse Detectors}

\newcommand\tud[0]{\textsuperscript{\normalfont \textdagger}}
\newcommand\iowa[0]{\textsuperscript{\normalfont \textparagraph}}
\newcommand\utd[0]{\textsuperscript{\normalfont \ddag}}
\newcommand\ualberta[0]{\textsuperscript{\normalfont \textasteriskcentered}}

\author{Sven~Amann,
  Hoan~Anh~Nguyen,
  Sarah~Nadi,
  Tien~N.~Nguyen,
  and~Mira~Mezini,~\IEEEmembership{Members,~IEEE}%
\IEEEcompsocitemizethanks{\IEEEcompsocthanksitem S. Amann and M. Mezini are with Technische Universit\"at Darmstadt, Germany.\protect\\
E-mails: amann@st.informatik.tu-darmstadt.de, mezini@informatik.tu-darmstadt.de
\IEEEcompsocthanksitem H. A. Nguyen is with Iowa State University, Iowa, United States of America.\protect\\
E-mail: hoan@iastate.edu
\IEEEcompsocthanksitem S. Nadi is with University of Alberta, Canada.\protect\\
E-mail: nadi@ualberta.ca
\IEEEcompsocthanksitem T. N. Nguyen is with University of Texas-Dallas, Texas, United States of America.\protect\\
E-mail: tien.n.nguyen@utdallas.edu}% <-this % stops an unwanted space
\thanks{Manuscript received July 1, 2017; revised January 20, 2018; accepted for publication March 12, 2018.}}

% The paper headers
\markboth{Accepted for Publication in IEEE TRANSACTIONS ON SOFTWARE ENGINEERING, 2018}%
{Amann \MakeLowercase{\textit{et al.}}: A Systematic Evaluation of Static API-Misuse Detectors}
% The only time the second header will appear is for the odd numbered pages
% after the title page when using the twoside option.
% 
% *** Note that you probably will NOT want to include the author's ***
% *** name in the headers of peer review papers.                   ***
% You can use \ifCLASSOPTIONpeerreview for conditional compilation here if
% you desire.

% for Computer Society papers, we must declare the abstract and index terms
% PRIOR to the title within the \IEEEtitleabstractindextext IEEEtran
% command as these need to go into the title area created by \maketitle.
% As a general rule, do not put math, special symbols or citations
% in the abstract or keywords.
\IEEEtitleabstractindextext{%
\input{sections/abstract}

% Note that keywords are not normally used for peerreview papers.
}

\maketitle

% To allow for easy dual compilation without having to reenter the
% abstract/keywords data, the \IEEEtitleabstractindextext text will
% not be used in maketitle, but will appear (i.e., to be "transported")
% here as \IEEEdisplaynontitleabstractindextext when the compsoc
% or transmag modes are not selected <OR> if conference mode is selected
% - because all conference papers position the abstract like regular
% papers do.
\IEEEdisplaynontitleabstractindextext
% For peer review papers, you can put extra information on the cover
% page as needed:
% \ifCLASSOPTIONpeerreview
% \begin{center} \bfseries EDICS Category: 3-BBND \end{center}
% \fi
%
% For peerreview papers, this IEEEtran command inserts a page break and
% creates the second title. It will be ignored for other modes.
\IEEEpeerreviewmaketitle

\input{sections/introduction}
\input{sections/misuses}
\input{sections/survey}
\input{sections/benchmark}
\input{sections/pipeline}
\input{sections/results}
\input{sections/threats}
\input{sections/conclusions}
\input{sections/acknowledgements}

\balance
\bibliographystyle{IEEEtran}
%\bibliographystyle{abbrv}
%\citestyle{acmauthoryear}
%\setcitestyle{numbers,sort&compress}

\bibliography{references/references,references/urls}

% biography section
% 
% If you have an EPS/PDF photo (graphicx package needed) extra braces are
% needed around the contents of the optional argument to biography to prevent
% the LaTeX parser from getting confused when it sees the complicated
% \includegraphics command within an optional argument. (You could create
% your own custom macro containing the \includegraphics command to make things
% simpler here.)
%\begin{IEEEbiography}[{\includegraphics[width=1in,height=1.25in,clip,keepaspectratio]{mshell}}]{Michael Shell}
% or if you just want to reserve a space for a photo:

\begin{IEEEbiography}[{\includegraphics[width=1in,height=1.25in,clip,keepaspectratio]{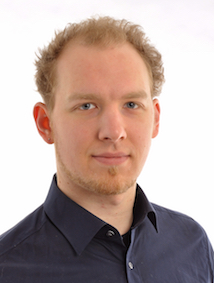}}]{Sven Amann}
is a doctoral candidate at TU Darmstadt, Germany. His primary research domain is API-misuse
detection using static analyses and machine-learning techniques applied to examples mined from
large code repositores and code search engines. Sven is founder and project lead of the MUBench
benchmark suite. More on
\ahref{https://web.archive.org/web/20170606231336/http://sven-amann.de}{http://sven-amann.de}.
\end{IEEEbiography}

\begin{IEEEbiography}[{\includegraphics[width=1in,height=1.25in,clip,keepaspectratio]{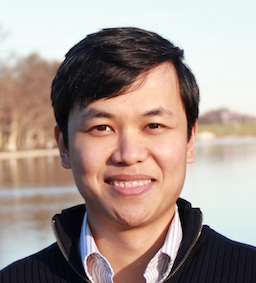}}]
{Hoan Nguyen}
is a post-doctoral researcher at Iowa State University. He received his PhD from Iowa State University. His research interests include program analysis, software evolution and maintenance, and mining software repositories.
\end{IEEEbiography}

\begin{IEEEbiography}[{\includegraphics[width=1in,height=1.25in,clip,keepaspectratio]{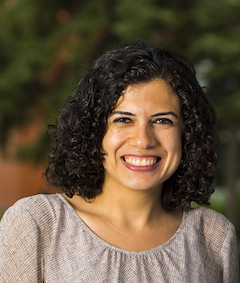}}]{Sarah Nadi}
is an Assistant Professor at the University of Alberta. She received both her MMath (2010) and PhD (2014) degrees from the University of Waterloo, Canada, and spent approximately two years as a post-doctoral researcher at TU Darmstadt, Germany. Her research interests include mining software repositories, software product lines, and helping developers use APIs correctly. 
\end{IEEEbiography}

\begin{IEEEbiography}[{\includegraphics[width=1in,height=1.25in,clip,keepaspectratio]{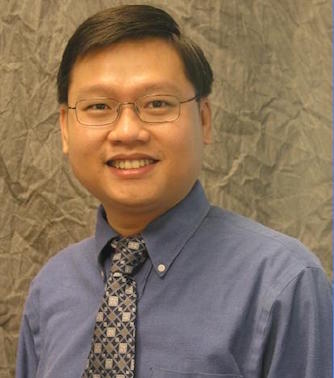}}]
{Tien Nguyen}
 is an associate professor at The University of Texas at Dallas. His research interests include program analysis, large-scale mining software repositories, and statistical approaches in software engineering. 
 \end{IEEEbiography}

\begin{IEEEbiography}[{\includegraphics[width=1in,height=1.25in,clip,keepaspectratio]{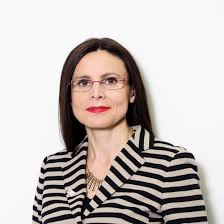}}]{Mira Mezini}
received the diploma degree in computer science from the
  University of Tirana, Albania, and the PhD degree in computer
  science from the University of Siegen, Germany.  She is a professor
  of computer science at the Tech\-ni\-sche Universit\"{a}t Darmstadt,
  Germany, where she heads the Software Technology Lab.
\end{IEEEbiography}

% insert where needed to balance the two columns on the last page with
% biographies
%\newpage

% You can push biographies down or up by placing
% a \vfill before or after them. The appropriate
% use of \vfill depends on what kind of text is
% on the last page and whether or not the columns
% are being equalized.

%\vfill

% Can be used to pull up biographies so that the bottom of the last one
% is flush with the other column.
%\enlargethispage{-5in}

\end{document}

%% file: meta/packages.tex
\usepackage[utf8]{inputenc}
\usepackage[ngerman,english]{babel}

\usepackage{epstopdf}

\usepackage{mathptmx}
\usepackage{microtype}

\usepackage{url}
\usepackage[hidelinks]{hyperref}

\usepackage{graphicx}
\usepackage[table,x11names]{xcolor}

\usepackage{xspace}

\usepackage{cite}
\usepackage{caption}
\usepackage{subcaption}

\usepackage{booktabs,colortbl}
\usepackage{adjustbox}
\newcolumntype{R}[2]{%
    >{\adjustbox{angle=#1,lap=\width-(#2)}\bgroup}%
    l%
    <{\egroup}%
}
\newcommand*\rotdia{\multicolumn{1}{R{45}{1em}}}% no optional argument here, please!
\newcommand*\rot{\rotatebox{90}}
\usepackage{multicol,multirow}

\usepackage{amsmath}
\usepackage{wasysym}
\usepackage{comment}
\usepackage{siunitx}

\usepackage[inline]{enumitem}
\usepackage{textcomp}
\usepackage{balance}

\input{meta/listings}
\input{meta/yaml-listings}

%% file: meta/listings.tex
\usepackage{listings}
\definecolor{KWColor}{rgb}{0.37,0.08,0.25}
\definecolor{CommentColor}{rgb}{0.133,0.545,0.133}
\definecolor{StringColor}{rgb}{0,0.126,0.941}

%% file: meta/yaml-listings.tex
\newcommand\YAMLcolonstyle{\color{red}\ttfamily}
\newcommand\YAMLkeystyle{\color{black}\ttfamily}
\newcommand\YAMLvaluestyle{\color{blue}\ttfamily}

\makeatletter

\newcommand\language@yaml{yaml}

\expandafter\expandafter\expandafter\lstdefinelanguage
\expandafter{\language@yaml}
{
  keywords={true,false,null,y,n},
  keywordstyle=\color{darkgray}\bfseries,
  basicstyle=\YAMLkeystyle\footnotesize,                                 % assuming a key comes first
  sensitive=false,
  comment=[l]{\#},
  morecomment=[s]{/*}{*/},
  commentstyle=\color{purple}\ttfamily,
  stringstyle=\YAMLvaluestyle\ttfamily,
  moredelim=[l][\color{orange}]{\&},
  moredelim=[l][\color{magenta}]{*},
  moredelim=**[il][\YAMLcolonstyle{:}\YAMLvaluestyle]{:},   % switch to value style at :
  morestring=[b]',
  morestring=[b]",
  literate =    {---}{{\ProcessThreeDashes}}3
                {>}{{\textcolor{red}\textgreater}}1     
                {|}{{\textcolor{red}\textbar}}1 
                {\ -\ }{{\mdseries\ -\ }}3,
}

\lst@AddToHook{EveryLine}{\ifx\lst@language\language@yaml\YAMLkeystyle\fi}
\makeatother

\newcommand\ProcessThreeDashes{\llap{\color{cyan}\mdseries-{-}-}}

%% file: meta/macros.tex
\input{meta/observations}

\newcommand{\checkNum}[1]{#1}

\newcommand{\ahref}[2]{\href{#1}{\nolinkurl{#2}}}

\newcommand{\code}[1]{{\texttt{\footnotesize{#1}}\xspace}}
\newcommand{\etal}{{\em et al.}}

\newcommand\InlineSec[1]{\vspace{0.03in}\noindent\textbf{#1.}}

\newcommand\aName[1]{{\small\textsc{#1}}\xspace}

\newcommand\MUBench[0]{\aName{MuBench}}
\newcommand\MUPipe[0]{\aName{MuBenchPipe}}
\newcommand\MUC[0]{\aName{MuC}}

\newcommand\GROUM[0]{\aName{GROUM}}

\newcommand\Colibri[0]{\aName{Colibri/ML}}
\newcommand\DroidAssist[0]{\aName{DroidAssist}}

\newcommand\GROUMiner[0]{\aName{GROUMiner}}
\newcommand\Jadet[0]{\aName{Jadet}}
\newcommand\PRMiner[0]{\aName{PR-Miner}}
\newcommand\CARMiner[0]{\aName{CAR-Miner}}
\newcommand\Alattin[0]{\aName{Alattin}}
\newcommand\Tikanga[0]{\aName{Tikanga}}
\newcommand\DMMC[0]{\aName{DMMC}}
\newcommand\RGJ[0]{\aName{RGJ07}}
\newcommand\Chronicler[0]{\aName{Chronicler}}
\newcommand\AX[0]{\aName{AX09}}

%% file: meta/observations.tex
\usepackage[absolute,overlay]{textpos}
\usepackage{tikz}
\usepackage{xcolor}
\usepackage{calc}

\newcounter{obsctr}
\newcommand{\obsref}[1]{\emph{O\ref{obs:#1}}}
\newcommand{\observation}[2]{\refstepcounter{obsctr}\emph{O\theobsctr:\label{obs:#1}}}

\newlength{\boxw}
\newlength{\boxh}
\newlength{\shadowsize}
\newlength{\boxroundness}
\newlength{\tmpa}
\newsavebox{\shadowblockbox}

\newenvironment{obs}[1]%
{\vspace{0.2cm}\noindent
\begin{lrbox}{
\shadowblockbox
}
\begin{minipage}{.98\columnwidth}
\observation{#1}~}%
{\end{minipage}\end{lrbox}%
\settowidth{\boxw}{\usebox{\shadowblockbox}}%
\settodepth{\tmpa}{\usebox{\shadowblockbox}}%
\settoheight{\boxh}{\usebox{\shadowblockbox}}%
\addtolength{\boxh}{\tmpa}%
\begin{tikzpicture}
\addtolength{\boxw}{\boxroundness * 2}
\addtolength{\boxh}{\boxroundness * 2}

\foreach \x in {0,.05,...,1}
{
\setlength{\tmpa}{\shadowsize * \real{\x}}
\fill[xshift=\shadowsize - 1pt,yshift=-\shadowsize + 
1pt,black,opacity=.04,rounded corners=\boxroundness] 
(\tmpa, \tmpa) rectangle +(\boxw - \tmpa - \tmpa, \boxh - \tmpa - 
\tmpa);
}

\filldraw[fill=white!50, draw=black!80, rounded corners=\boxroundness] (0, 
0) rectangle (\boxw, \boxh);
\draw node[xshift=\boxroundness,yshift=\boxroundness,inner sep=0pt,outer 
sep=0pt,anchor=south west] (0,0) {\usebox{\shadowblockbox}};
\end{tikzpicture}\vspace{0cm}%
}

%% file: sections/abstract.tex
%!TEX root = ../paper.tex

\begin{abstract}
Application Programming Interfaces (APIs) often have usage constraints, such as restrictions on call order or call conditions.
\textit{API misuses}, i.e., violations of these constraints, may lead to software crashes, bugs, and vulnerabilities.
Though researchers developed many API-misuse detectors over the last two decades, recent studies show that API misuses are still prevalent.
Therefore, we need to understand the capabilities and limitations of existing detectors in order to advance the state of the art.
In this paper, we present the first-ever qualitative and quantitative evaluation that compares static API-misuse detectors along the same dimensions, and with original author validation.
To accomplish this, we develop \MUC, a classification of API misuses, and \MUPipe, an automated benchmark for detector comparison, on top of our misuse dataset, \MUBench.
Our results show that the capabilities of existing detectors vary greatly and that existing detectors, though capable of detecting misuses, suffer from extremely low precision and recall.
A systematic root-cause analysis reveals that, most importantly, detectors need to go beyond the naive assumption that a deviation from the most-frequent usage corresponds to a misuse and need to obtain additional usage examples to train their models.
We present possible directions towards more-powerful API-misuse detectors.
\end{abstract}

%% file: sections/introduction.tex
%!TEX root = ../paper.tex

\IEEEraisesectionheading{\section{Introduction} % (fold) 
\label{sec:introduction}}

% Motivation

\IEEEPARstart{I}{ncorrect} usages of an Application Programming Interface (API), or \emph{API misuses}, are violations of (implicit) \emph{usage constraints} of the API. 
An example of a usage constraint is having to check that \code{hasNext()} returns \code{true} before calling \code{next()} on an \code{Iterator}, in order to avoid a \code{NoSuchElementException} at runtime.
Incorrect usage of APIs is a prevalent cause of software bugs, crashes, and vulnerabilities~\cite{MM13,SHA15,ANNN+16,FHMB+12,EBFK13,NKMB16,GIJA+12}.
While high-quality documentation of an API's usage constraints could help, it is often insufficient, at least in its current form, to solve the problem~\cite{DH09}.
For example, a recent empirical study shows that Android developers prefer informal references, such as StackOverflow, over official API documentation, even though the former promotes many insecure API usages~\cite{ABFKMS16}.
We confirm this tendency for a non-security API as well:
Instances of \code{Iterator} may not be used after the underlying collection was modified, otherwise they throw a \code{ConcurrentModificationException}.
Even though this constraint and the consequences of its violation are thoroughly documented, a review of the \checkNum{top-5\%} of \checkNum{2,854} threads about \code{ConcurrentModificationException} on StackOverflow shows that \checkNum{57\%} of them ask for a fix of the above misuse~\cite{artifact-page}.

Ideally, development environments should assist developers in implementing correct usages and in finding and fixing existing misuses.
In this paper, we focus on tools that identify misuses in a given codebase, specifically, those that automatically infer~API-usage specifications and identify respective violations through static code analysis.
We refer to these tools as \emph{static API-misuse~detectors}.

% What happened before...
There have been many attempts to address the problem of API misuse.
Existing static misuse detectors commonly mine \textit{usage patterns}, i.e., equivalent API usages that occur frequently, and report any anomaly with respect to these patterns as potential misuse~\cite{LZ05,L07,WZL07,RGJ07,NNP+09,AX09,TX09,TX09b,WZ11,MM13,NPVN16}.
The approaches differ in how they encode usages and frequency, as well as in the techniques they apply to identify patterns and violations thereof.
Despite the vast amount of work on API-misuse detection, API misuses still exist in practice, as recent studies show~\cite{LHXRM16,ABFKMS16}.
To advance the state of the art in API-misuse detection, we need to understand how existing approaches compare to each other, and what their current limitations are.
This would allow researchers to improve API-misuse detectors by enhancing current strengths and overcoming weaknesses.

%We need a classification of API-misuses to assess each detector's capabilities and to systematically address the problem space.
%And we need means to empirically evaluate and compare detectors.

In this work, we propose the \textit{API-Misuse Classification} (\MUC) as a taxonomy for API misuses and a framework to assess the capabilities of static API-misuse detectors.
In order to create such a taxonomy, we need a diverse sample of API misuses.
In our previous work, we described \MUBench, a dataset of \checkNum{90 API misuses} that we collected by reviewing \checkNum{over 1200 reports from existing bug datasets} and conducting a developer survey~\cite{ANNN+16}.
\MUBench provided us with the misuse examples needed to create a taxonomy.
To cover the entire problem space of API misuses, for this paper, we add further misuses to this dataset by looking at examples from studies on API-usage directives~\cite{DH09,METM12}. 
Using \MUC, we qualitatively compare \checkNum{12 existing detectors} and identify their shortcomings.
For example, we find that only few detectors detect misuses related to conditions or exception handling.
We confirm this assessment with the detectors' original authors.

The previous step provides us with a conceptual comparison of existing detectors.
We also want to compare these API-misuse detectors empirically, by both their precision and recall.
This is a challenging task, due to the different underlying mechanisms and representations used by detectors.
To enable this empirical comparison, we build \MUPipe, the first automated pipeline to benchmark API-misuse detectors.
Our automated benchmark leverages \MUBench, and the additional misuses we collect in this work, and creates an infrastructure on top of it to run the detectors and compare their results.
We perform three experiments based on \checkNum{29 real-world projects} and \checkNum{25 hand-crafted examples} to empirically evaluate and compare \checkNum{four state-of-the-art detectors}.
We exclude the \checkNum{other eight detectors} since \checkNum{two} rely on the discontinued Google Code Search~\cite{farewellGoogle}, \checkNum{five} target C/C++ code, and \checkNum{one} targets Dalvik Bytecode, while our benchmark contains Java misuses. 
In \nameref{e2}, we measure the precision of the detectors in a per-project setup, where they mine patterns and detect violations in individual projects from \MUBench.
In \nameref{e1}, we determine upper bounds to the recall of the detectors with respect to the known misuses in \MUBench.
We take the possibility of insufficient training data out of the equation, by providing the detectors with crafted examples of correct usages for them to mine required patterns.
Finally, in \nameref{e3}, we measure the recall of the detectors against both the \MUBench dataset and the detectors' own confirmed findings from \nameref{e2} using a per-project setup.

Our conceptual analysis shows many previously neglected aspects of API misuse, such as incorrect exception handling and redundant calls.
Our quantitative results show that misuse detectors are capable of detecting misuses, when provided with correct usages for pattern mining.
However, they suffer from extremely low precision and recall in a realistic setting. 
We identify \checkNum{four root causes} for false negatives and \checkNum{seven root causes} for false positives. 
Most importantly, to improve precision, detectors need to go beyond the naive assumption that a deviation from the most-frequent usage corresponds to a misuse, for example, by building probabilistic models to reason about the likelihood of usages in their respective context.
To improve recall, detectors need to obtain more correct usage examples, possibly from different sources, and to consider program semantics, such as type hierarchies and implicit dependencies between API usages.
These novel insights are made possible by our automated benchmark.
Our empirical results present a wake-up call, unveiling serious practical limitations of tools and evaluation strategies from the field.
Foremost, detectors suffer from extremely low recall---which is typically not evaluated. 
Moreover, we find that the application of detectors to individual projects does not seem to give them sufficient data to learn good models of correct API usage.

%For the first time, we systematically measure the recall of detectors and unveil their poor performance.
%This should call researchers to action; we provide our tooling and data to enable replication of experiments for additional detectors.

%Experiment_3 shows that all detectors have very low recall, likely because they cannot mine the required patterns from within single projects.
%This calls for new cross-project mining approaches.

% Contribution

In summary, this paper makes the following contributions to the area of API-misuse detection:
\begin{itemize}
  \item A taxonomy of API misuses, \MUC, which provides a conceptual framework to compare the capabilities of API-misuse detectors.
  \item A survey and qualitative assessment of \checkNum{12 state-of-the-art} misuse detectors, based on \MUC.
  \item A publicly available automated benchmark pipeline for API-misuse detectors, \MUPipe, which facilitates systematic and reproducible evaluations of misuse detectors.
  \item An empirical comparison of both recall and precision of~\checkNum{four existing misuse detectors} using \MUPipe.
  Our work is the first to compare different detectors on both a conceptual and practical level and, more importantly, the first to measure the recall of detectors, unveiling their poor performance.
  \item A systematic analysis of the root causes for low precision and recall across detectors, to call researchers to action.
\end{itemize}

Our benchmarking infrastructure is publicly available~\cite{mubench} and our artifact Web page~\cite{artifact-page} provides full details on our results.

%\todo{Use formulations from rebuttal to strengthen introduction?}

% How this is awesome, saves lifes, and advances the state-of-the-art

% section introduction (end)

%% file: sections/misuses.tex
%!TEX root = ../paper.tex

\section{Background and Terminology}
\label{sec:definitions}

An \emph{API usage} (\textit{usage}, for short) is a piece of code that uses a given API to accomplish some task.
It is a combination of basic~\textit{program elements}, such as method calls, exception handling, or arithmetic operations.
The combination of such elements in an API usage is subject to constraints, which depend on the nature of the API.
We call such constraints \emph{usage constraints}.
For example, two methods may need to be called in a specific order, division may not be used with a divisor of zero, and a file resource needs to be released along all execution paths.
When a usage violates one or more such constraints, we call it a \emph{misuse}, otherwise a \emph{correct usage}. 

The detection of API misuses may be approached through \emph{static analyses} of source code or binaries and through \emph{dynamic analyses}, i.e., runtime monitoring or analysis of runtime data, such as traces or logs.
In either case, the detection requires either specifications of correct API usage to find violations of or specifications of misuses to find instances of.
Such specifications may be \emph{crafted manually} by experts or \emph{inferred automatically} by algorithms.
Automatic specification inference (or \textit{mining}) may, again, be approached both \emph{statically}, e.g., based on code samples or documentation, and \emph{dynamically}, e.g., based on traces or logs.

Since manually crafting and maintaining specifications is costly, in this work, we focus on automated detectors.
We call such tools \emph{API-misuse detectors}.
In the literature, we find \emph{static misuse detectors}, which statically mine specifications and detect misuses through static analysis, e.g., \cite{WZL07,NNP+09,MM13};
\emph{dynamic misuse detectors}, which dynamically mine specifications and detect misuses through dynamic analysis, e.g., \cite{PG12,LZLJMSR14};
and \emph{hybrid misuse detectors}, which, for example, combine dynamic specification mining with static detection~\cite{PJAG12}.
In this work, we focus on static API-misuse detectors.
 
Static API-misuse detection is often achieved through detecting \emph{deviant code}~\cite{ECHC+01,LZ05,L07,WZL07,RGJ07,NNP+09,AX09,TX09,TX09b,WZ11,MM13,NPVN16}.
The key idea is that mistakes violate constraints that the code should adhere to and that, given sufficiently many examples of correct usage, such violations appear as \emph{anomalies}. 
We call a usage that appears frequently in programs a \emph{pattern}.
The identification of mistakes through the detection of deviant code assumes that patterns correspond to correct usages (specifications) and anomalies with respect to these patterns are, consequently, misuses.
Such an approach can detect mistakes in the usage of popular libraries~\cite{ECHC+01,LZ05,WZL07,NNP+09,WZ11,MM13}.

In our previous work~\cite{ANNN+16}, we collected a dataset of Java API misuses by reviewing bug reports of \checkNum{21 real-world projects} and surveying developers about API misuses.
We call this dataset \MUBench.
It contains \checkNum{90 misuses}, \checkNum{73 misuses} from the real-world projects and \checkNum{17} from the survey (see \autoref{tab:datasets}, Row 1).
For each real-world misuse, the dataset identifies the \emph{project} where the misuse is, the \emph{project version} that contains the misuse, and the commit that fixed the misuse.
For the other misuses, \MUBench provides hand-crafted misuse examples and their fixes.

\begin{table}[tb]
  \centering
  \input{sections/table-datasets}
  \label{tab:datasets}
\end{table}

\section{The API-Misuse Classification (\MUC)} % (fold)
\label{sec:api_misuses}

In this section, we introduce the  \emph{API-Misuse Classification} (\MUC), our taxonomy for API misuses.
We derive \MUC from the misuse examples in the \MUBench dataset.
In \autoref{sec:misuse-detectors-related-work}, we use \MUC to qualitatively compare the capabilities of existing API-misuse detectors.
In \autoref{sec:results}, we use \MUC to define our expectations on the detectors' performance.
Before presenting the classification itself, we briefly~discuss existing related classifications to motivate the need for \MUC.

\subsection{Motivation for \MUC}
%% Classifications

IEEE has a standard for classifying software defects~\cite{IEEE10}, which served as the basis for IBM’s \aName{Orthogonal Defect Classification} (ODC)~\cite{CBCHMRW92}.
The ODC uses the defect type as one of the aspects from which to classify defects.
The defect type is composed of a conceptual program element, such as a function, check, assignment, documentation, or algorithm, and a violation type, i.e., either \emph{missing} or \emph{incorrect}.
More recently, Beller~\etal~\cite{BBMZ16} presented the \aName{General Defect Classification} (GCD), a remote ODC-descendant, tailored to compare the capabilities of automated static-analysis tools.
Both classifications capture the entire domain of all types of software defects.
To compare the capabilities of API-misuse detectors, we need a more fine-grained differentiation of a subset of the categories in both of them.

Past work presented empirical studies and taxonomies of API-usage directives~\cite{DH09,METM12}.
Many of these directives can be thought of as usage constraints in our terminology and their violations, consequently, as misuses.
Other directives, however, do not formulate constraints.
Examples are directives that explicitly allow \code{null} to be passed as a parameter and directives that inform about alternative ways to achieve a behaviour (possibly with different trade-offs).
Therefore, we cannot directly convert a taxonomy of usage directives into a taxonomy of misuses.
Instead, to consider the directives that can be viewed as usage constraints, we extend \MUBench~\cite{ANNN+16} by hand-crafted examples of misuses violating them, which we derive from examples in the studies.
This gives us \checkNum{10 additional misuses}, resulting in a total of \checkNum{100 misuses} that we use for \MUC and our experiments (see \autoref{tab:datasets}, Row 2).
For simplicity, we subsequently refer to this extended dataset as \MUBench.

%In the \MUBench data paper~\cite{ANNN+16}, we provided a basic API-misuse classification.
%However, in this work, we want to provide a more comprehensive and detailed classification to enable us to compare existing misuse detectors.
%In addition to the original \MUBench misuses, we consider the findings of two empirical studies on API usage directives~\cite{DH09,METM12}.
%Many of the directives those studies identify can be thought of as usage constraints in our terminology and their violation, consequently, as misuses.
%The studies report directives corresponding to all violations in \MUBench and a few more.~\sn{doesn't that make it sound like we could have just used these directives since they correspond to exactly what we have plus more?}
%To make \MUBench more comprehensive, we add hand-crafted examples of misuses violating these additional directives, which we derive from the examples in the studies.

\subsection{The Classification}

We developed \MUC using a variation of Grounded Theory~\cite{GS67}:
Following our notion of API misuses as API usages with one or more violations of usage constraints, the first author of this work went through all the misuses in \MUBench and came up with labels for the characteristics of the respective violations, until each misuse was tagged with at least on label.
Subsequently, all authors iteratively revisited the labelled misuses to unify semantically equivalent labels and group related labels, until we had a consistent taxonomy.
In the end, we had two dimensions whose intersection describes all violations in \MUBench:
the type of the involved API-usage element and the type of the violation.
Consequently, we define a \emph{violation} as a pair of a violation type and an API-usage element.

An \textit{API-usage element} is a program element that appears in API usages.
The following elements are involved in the misuses in \MUBench:
\emph{method calls}, \emph{conditions}, \emph{iterations}, and \emph{exception handling}.
Note that we consider primitive operators, such as arithmetic operators, as methods.
For conditions, we further distinguish \emph{\code{null} checks}, \emph{value or state conditions}, \emph{synchronization conditions}, and \emph{context conditions}, because of their distinct properties.

The \textit{violation type} describes how a usage violates a given usage constraint with respect to a given usage element.
In \MUBench, we find \checkNum{two} violation types: {\em missing} and {\em redundant}.
Violations of the missing type come from constraints that mandate the presence of a usage element.
They generally cause program errors.
An example of such a violation is a ``missing method call.''
Violations of the redundant type come from constraints that mandate the absence of a usage element or declare the presence of a usage element unnecessary.
Note that in either case the repetition of an element may have undesired effects, such as errors or decreased performance.
An example of such a redundant violation is a ``redundant method call.''

\autoref{tab:muc} shows a summary of \MUC.
The numbers in the cells show how many misuses in \MUBench have a respective violation.
Note that a single misuse may have multiple violations; thus, the individual cells in the table sum up to more than \checkNum{100}.
The table shows that missing method calls, \code{null} checks, and value or state conditions are the most prevalent violations.
Redundant calls and missing exception handling are less frequent, but still prevalent, while we have only few examples of the other violations.

We now discuss the different violation categories shown in \autoref{tab:muc}, grouped by the API-usage element involved.

\begin{table}
  \centering
  \caption{The Misuse Classification (\MUC), with the number of misuses with a particular violation in \MUBench.}
    \input{sections/table-muc}
  \label{tab:muc}
\end{table}

\subsubsection*{Method Calls}

Method calls are the most prominent elements of API usages, as they are the primary means of communication between client code and the API.

One violation category is \emph{missing method calls}, which occur if a usage does not call a certain method that is mandated by the API usage constraints.
For example, if a usage does not call \code{validate()} on a \code{JFrame} after adding elements to it, which is required for the change to become visible.

The other case is \emph{redundant method calls}, which occur if a usage calls a certain method that is restricted by the API usage constraints.
For example, if a usage calls \code{remove()} on a list that is currently being iterated over, which causes an exception in the subsequent iteration.
% or when \code{finalize()} is called on an \code{Object}, which should never be done from any user-defined code.

\subsubsection*{Conditions}

Client code often needs to ensure conditions for valid communication to an API, in order to adhere to the API's usage constraints.
There are often alternative ways to ensure such conditions.
For example, to ensure that a collection is not empty one may check \code{isEmpty()}, check its \code{size()}, or add an element to it.
Note that checks, in particular, are also a means for the client code to vary usages depending on program inputs.

One violation category is \emph{missing conditions}, which occur if a usage does not ensure certain conditions that are mandated by the API usage constraints.
One case is \emph{missing null checks}, e.g., if a usage fails to ensure that a receiver or a parameter of a call is not \code{null}.
Another case is \emph{missing value or state conditions}, e.g., if a usage fails to ensure that a \code{Map} contains a certain key before using the key to access the \code{Map}.
In multi-threaded environments, \emph{missing synchronization conditions} may occur, e.g., if a usage does not obtain a lock before updating a \code{HashMap} that is accessed from multiple threads~\cite{METM12}.
Finally, \emph{missing context conditions} may also occur, e.g., if a usage fails to ensure that GUI components in \aName{Swing} are updated on the Event Dispatching Thread (EDT)~\cite{DH09}.
% and \code{Logger.setParent()} should not be called from application code~\cite{METM12}.

The other case is \emph{redundant conditions}, where a condition prevents a necessary part of a usage, e.g., a method call, from being executed along certain execution paths or is simply redundant.
One case is \emph{redundant \code{null} checks}, e.g., if the usage checks nullness only after a method has been invoked on the respective object.
Another case is \emph{redundant value or state conditions}, e.g., if the usage checks \code{isEmpty} on a collection that's guaranteed to contain an element.
In multi-thread environments, \emph{redundant synchronization conditions} may occur, e.g., if the usage requests a lock that it already holds, which may cause a deadlock.
Finally, \emph{redundant context conditions} may also occur, e.g., if a \aName{JUnit} assertion is executed on another thread, where its failing cannot be captured by the \aName{JUnit} framework.

\subsubsection*{Iteration}

Iteration is another means of interacting with APIs, used, in particular, with collections and IO streams.
It takes the form of loops and recursive methods.
Note that respective usage constraints are about (not) repeating (part of) a usage, rather than about the condition that controls the execution.
%~\mm{The Iteration category doesn't feels right to me. Aren't the two examples given rather instances of other misusages? The first one could be seen as "check before wait" and the loop is simply a way to ensure this constraint, while the second one can be seen as a superflous call ...
% Where do they come from? And, given that there are only two, does it make sense to get rid of them altogether?}
% \hn{``check before wait'' would over-simplify the usage where wait() is kept being called until the condition is no longer satisfied. The second one cannot be classified as a redundant call be cause the call must be there. } 

One violation category is \emph{missing iterations}, which occur if a usage does not repeatedly check a condition that the API usage constraints mandate must be checked again after executing part of the usage.
For example, the Java documentation states that a call to \code{wait()} on an object should always happen in a loop that checks the condition the code waits for, because \code{wait()} could return before the condition is satisfied, in which case the usage should continue to wait.
%Note that even if the code ensures the correct condition, say, with an \code{if}, the iteration constraint is still violated since the check and, depending on its outcome, the invocation of \code{wait()} are not repeated.\hn{The last sentence is not needed and could confuse readers. The previous sentence already says ``should always happen in a loop''}

The other case is \emph{redundant iterations}, which occur if part of a usage is reiterated that the API usage constraints mandate may be executed not more than once or that is simply redundant.
For example, a \code{Cipher} instance might be reused in a loop to encrypt a collection of values, but its initialization through calling \code{init()} must happen exactly once, i.e., before the loop.
Note that in this situation, the required call is present in the respective code exactly once, as required, but its inclusion in an iteration causes a violation.

\subsubsection*{Exception Handling}

Exceptions are a way for APIs to communicate errors to client code.
The handling of different errors often depends on the specific~API.

One violation category is \emph{missing exception handling}, which occurs if a usage does not take actions to recover from a possible error, as mandated by the API usage constraints.
For example, when initializing a \code{Cipher} with an externally provided cryptographic key, one should handle \code{InvalidKey\-Exception}.
Another example is resources that need to be closed after use, also in case of an exception.
Such guarantees are often implemented by a \code{finally} block, but also using the try-with-resources construct or even respective handling in multiple \code{catch} blocks.

The other case is \emph{redundant exception handling}, which occurs if a usage intercepts exceptions that should not be caught or handled explicitly.
For example, catching \code{Throwable} when executing a command in an application might suppress a \code{CancellationException}, preventing the user from cancelling the execution.

%% file: sections/table-datasets.tex
%!TEX root = ../paper.tex

\caption{Datasets used throughout this paper, with the number of hand-crafted misuses (\#HM), the number of real-world projects (\#P), project versions (\#PV), and misuses (\#RM), and the total number of misuses (\#M).
``n/a'' denotes that the number is not relevant for the use of the dataset.}
\begin{tabular}{rlrrrrr}
  \toprule
   & Dataset & \#HM & \#P & \#PV & \#RM & \#M \\
  \midrule
  1 & Original \MUBench~\cite{ANNN+16} &  17 & 21 & 55 &  73 &  90 \\
  2 & Extended \MUBench                &  27 & 21 & 55 &  73 & 100 \\
  3 & \nameref{e2}                     & n/a &  5 &  5 & n/a & n/a \\
  4 & \nameref{e1}                     &  25 & 13 & 29 &  39 &  64 \\
  5 & \nameref{e3}                     &   0 & 13 & 29 &  53 &  53 \\
  \bottomrule
\end{tabular}

%% file: sections/table-muc.tex
%!TEX root = ../paper.tex

\begin{tabular}{lrr}
  \toprule
  & \multicolumn{2}{c}{Violation Type}\\
  \cmidrule{2-3}
  API-Usage Element & Missing & Redundant \\
  \midrule
  Method Call                    & 30 &      13 \\
  Condition                      & 48 &       6 \\
  \hspace{1em}\code{null} Check & 25 &       3 \\
  \hspace{1em}Value or State    & 21 &       2 \\
  \hspace{1em}Synchronization   &  1 &       1 \\
  \hspace{1em}Context           &  1 &       1 \\
  Iteration                     &  1 &       1 \\
  Exception Handling            & 10 &       1 \\
  \bottomrule
\end{tabular}

%% file: sections/survey.tex
%!TEX root = ../paper.tex

\section{Conceptual Classification of Existing Misuse Detectors}
\label{sec:misuse-detectors-related-work}

To advance the state of the art of API-misuse detection, we need to understand the capabilities and short-comings of existing misuse detectors.
To identify detectors, we started from the publications about API-misuse detection listed in a survey of automated API-property inference techniques by Robillard~\etal~\cite{RBMR13}.
For each publication, we looked at all publications they refer to as related work and all publication that cite them, according to the ACM Digital Library or the IEEE Xplore Digital Library.
We recursively repeated this process, until we found no new detectors.

We use \MUC to guide the comparison.
We provide a \textit{conceptual classification} of the \emph{capabilities} of each detector with respect to \MUC, summarized in \autoref{tab:detector_capabilities}.
We use the published description and results of each detector to identify which of \MUC categories they can, conceptually, detect.
To reduce subjectivity, we confirmed our capability assessment and the detector descriptions with the respective authors, except for \PRMiner and \Colibri, whose authors did not respond. 
We also describe the strategies used to evaluate each detector and summarize those in \autoref{tab:survey-evaluations}.

% Unless otherwise stated, the surveyed detectors perform both pattern mining and anomaly detection at the same time (i.e., not as two separate phases), and all evaluations focused on detecting project-specific violations.
% This means that patterns mined from a specific project are used to detect misuses in the same project.

% summarize the evaluation methodology and results of each detector in \autoref{tab:survey_evaluation}.
% Column shows the number of evaluated projects.
% Most authors followed a ``Top X'' approach for selecting the anomalies they review (Column 2), where X could be a fixed number or percentage.
% Column 3 shows the true positive rate of reviewed anomalies. 

\vspace{0.03in}

%\paragraph*{\PRMiner}
\PRMiner is a misuse detector for C~\cite{LZ05}.
It encodes usages as the set of all function names called within the same function and then employs frequent-itemset mining to find patterns with a minimum support of $15$ usages.
% Detection
Violations here are strict subsets of a pattern that occur at least ten times less frequently than the pattern.
To prune false positives, \PRMiner{} applies inter-procedural analysis, i.e., for each occurrence of a violation, it checks whether the missing call occurs within a called method.
This analysis follows the call path for up to 3 levels.
The reported violations are ranked by the respective pattern's support.
% Capabilities
\PRMiner focuses on detecting missing method-calls.
% Performance
The evaluation applied \PRMiner to three target projects individually, thereby finding violations of project-specific patterns.
The detector reported 1,601 findings (1,447, 147, and 7 on the individual projects).
The authors reviewed the top-60 violations reported across all projects and found 18.1\% true positives (26.7\%, 10.0\%, and 14.3\% on the individual projects).

\vspace{0.03in}

\Chronicler is a misuse detector for C~\cite{RGJ07b}.
It mines frequent call-precedence relations from an inter-procedural control-flow graph.
A relation is considered frequent, if it holds on at least 80\% of all execution paths.
Paths where such relations do not hold are reported as violations.
% Capabilities
\Chronicler detects missing method calls.
Since loops are unrolled exactly once, it cannot detect missing iterations.
% Performance
The evaluation applied \Chronicler to five projects, thereby finding violations of project-specific patterns.
The authors compared the identified protocols with the documented protocols for one API and discussed a few examples of actual bugs found by their tool.

\vspace{0.03in}

%\paragraph*{\Colibri{}}
\Colibri is another misuse detector for C~\cite{L07}.
It re-implements \PRMiner using \emph{Formal Concept Analysis}~\cite{GW99} to strengthen the theoretical foundation of the approach.
% Capabilities
Consequently, its capabilities are the same as \PRMiner's.
% Performance
The evaluation applied \Colibri to five target projects, thereby finding violations of project-specific patterns.
While some detected violations are presented in the paper, no statistics on the quality of the detector's findings are reported.

\vspace{0.03in}

\input{sections/table-capabilities}

\input{sections/table-survey-evaluation}

%\paragraph*{\Jadet}~\cite{WZL07}
\Jadet is a misuse detector for Java~\cite{WZL07}.
It uses \Colibri{}~\cite{L07}, but instead of only method names, it encodes method-call order and call receivers in usages.
It builds a directed graph whose nodes represent method calls on a given object and whose edges represent control flows.
From this graph, it derives a pair of calls for each call-order relationship.
%, e.g., \code{m() $\prec$ n()}. 
The sets of these pairs form the input to the mining, which identifies patterns, i.e., sets of pairs, with a minimum support of $20$.
% Detection
A violation may miss at most $2$ properties of the violated pattern and needs to occur at least ten times less frequently than the pattern.
Detected violations are ranked by $u \times s / v$, where $s$ is the violated pattern's support, $v$ is the number of violations of the pattern, and $u$ is a uniqueness factor of the pattern.
% Capabilities
\Jadet detects missing method calls.
It may detect missing loops as a missing call-order relation from a method call in the loop header to itself.
%However, it cannot detect violations of patterns that consist of only two calls since such a pattern would be encoded as a set of a single pair of method calls.
% The only strict subset of such a pattern is the empty set, which is by definition not a violation.
% Performance
The evaluation applied \Jadet to five target projects, thereby finding violations of project-specific patterns.
The authors reviewed the top-10 violations reported per project and found 6.5\% true positives (0\%, 0\%, 7.7\%, 10.5\%, and 13.3\% on the individual projects).
Other findings were classified as code smells (6.5\%) or hints (35.0\%).

In a subsequent study, \Jadet was applied in a cross-project setting where it was applied to 6,097 projects at once, using a minimum pattern support of 200~\cite{GWZ10}.
The authors reviewed the top-25\% findings from a random sample of 20 projects, a total of 50 findings, and found 8\% true positives.
Other findings were classified as code smells (14.0\%).

\vspace{0.03in}

%\paragraph*{\RGJ}~\cite{RGJ07}
\RGJ is a misuse detector for C~\cite{RGJ07}.
It encodes usages as sets of properties for each variable. % \code{v}.
Properties are
comparisons to literals,
%e.g., $(\neq, \code{null})$,
argument positions in function calls,
%e.g., $(\operatorname{arg}(2), \code{f})$ if \code{v} was passed as the second argument to a function \code{f},
and assignments.
%, e.g., $(:=, \operatorname{res}(\code{f}))$ if the \code{v} was assigned the result of a call to \code{f}.
For each call, it creates a group of the property sets of the call's arguments.
To all groups for a particular function, it applies sequence mining to learn common sequences of control-flow properties and frequent-itemset mining to identify all common sets of all other property types.
% Detection
Subsequently, it identifies violations of the common~property sequences and sets.
% Capabilities
\RGJ is designed to detect missing conditions.
From the properties it encodes, it can detect missing \code{null} checks and missing value or state conditions.
Since patterns contain preceding calls on arguments, it may also detect missing calls, if the respective call shares an argument with another call in the pattern.
% Performance
The evaluation applied \RGJ to a single project, thereby finding violations of project-specific patterns.
The authors discussed several examples of actual bugs their approach detects, but reported no statistics on the detection performance.

\vspace{0.03in}

%\paragraph*{\Alattin}~\cite{TX09b}
\Alattin is a misuse detector for Java~\cite{TX09b}, specialized in alternative patterns for condition checks.
For each target method \code{m}, it queries a code-search engine to find example usages.
From each example, it extracts a set of rules about pre- and post-condition checks on the receiver, the arguments, and the return value of \code{m}, e.g.,  ``\code{boolean} check on \code{return} of \code{Iterator.hasNext} before \code{Iterator.next}.''
% or ``\code{null} check on \code{argument} of \code{ArrayList.constructor} after \code{Iterator.next}.'' 
It then applies frequent-itemset mining on these rules to obtain patterns with a minimum support of $40\%$.
For each such pattern, it extracts the subset of all groups that do not adhere to the pattern and repeats mining on that subset to obtain infrequent patterns with a minimum support of $20\%$.
Finally, it combines all frequent and infrequent patterns for the same method by disjunction.
% Detection
An analyzed method has a violation if the set of rules that hold in it is not a superset of any of the alternative patterns.
Violations are ranked by the support of the respective pattern.
% Capabilities
\Alattin, therefore, detects missing \code{null}-checks and missing value or state conditions that are ensured by checks and do not involve literals.
It may also detect missing method-calls that occur in checks.
% Performance
The evaluation applied \Alattin to six projects.
Since it queries code-search engines for usage examples, it detects violations of cross-project patterns.
The authors manually reviewed all violations of the top-10 patterns per project, a total of 532 findings, and confirmed that 29.5\% identify missing condition checks (12.5\%, 26.2\%, 28.1\%, 32.7\%, 52.6\%, and 100\% for the individual projects).
Considering frequent alternative patterns reduced false positives by 15.2\% on average, which increased precision to 33.3\%.
Considering both frequent and infrequent alternatives even reduced false positives by 28.1\% on average, leading to a precision of 37.8\%, but introduced 1.5\% additional false negatives, because misuses that occur multiple times are mistaken for infrequent patterns.

\vspace{0.03in}

%\paragraph*{\AX}~\cite{AX09}
\AX is a misuse detector for C~\cite{AX09}, specialized in detecting wrong error handling, realized through returning (and checking for) error codes.
It distinguishes normal paths, i.e., execution paths from the beginning of the \code{main} function to its end, and error paths, i.e., paths from the beginning of the \code{main} function to an \code{exit} or \code{return} statement in an error-handling block.
\AX uses push-down model checking to generate such paths as sequences of method calls and applies frequent-subsequence mining to find patterns with a minimum support of 80\% (but at least 5 usages).
% Detection
It then uses push-down model checking to verify adherence to these patterns and identify respective violations.
Finally, it filters false positives by tracking variable values and excluding error cases that cannot occur.
% rank violations by pattern support
% Capabilities
It detects missing error-handling as well as missing method calls among error-handling functions.
%As a side effect of identifying
Since it identifies error-handling blocks through a predefined set of checks, it also detects missing \code{null}-checks and missing value or state conditions in the case of missing error-handling blocks.
% Performance
The evaluation applied \AX to three projects individually, thereby finding violations of project-specific patterns.
The authors manually reviewed all 292 findings and confirmed $90.4\%$ true positives (50.0\%, 90.3\%, and 93.5\% on the individual projects).

\vspace{0.03in}

%\paragraph*{\CARMiner}
\CARMiner is a misuse detector for C++ and Java~\cite{TX09}, also specialized in detecting wrong error handling.
For each analyzed method \code{m} in a given code corpus, it queries a code-search engine to find example usages.
From the examples, it builds an \emph{Exception Flow Graph} (EFG), i.e., a control-flow graph with additional edges for exceptional flow. 
%These graphs distinguish exception edges, i.e., edges to and within \code{catch} and \code{finally} blocks. 
From the EFG, it generates \emph{normal} call sequences that lead to the currently analyzed call and \emph{exception} call sequences that lead from the call along exceptional edges.
Subsequently, it mines association rules between normal sequences and exception sequences, with a minimum support of $40\%$.
% and ranks these rules by their support.
% Detection
To detect violations, \CARMiner extracts the normal call sequence and the exception call sequence for the target method call.
It then uses the learned association rules to determine the expected exception handling and reports a violation if the actual sequence does not include it.
% Capabilities
\CARMiner detects missing exception-handling as well as missing method calls among error-handling functions.
%Since it does not distinguish calls in \code{finally} blocks from calls in \code{catch} blocks, it cannot detect missing guarantees.
% Performance
The evaluation applied \CARMiner to five projects.
Since it queries code-search engines for usage examples, it detects violations of cross-project patterns.
The authors manually reviewed all violations of the top-10 association rules for each project, a total of 264 violations, and confirmed that 60.1\% identify wrong error handling (41.1\%, 54.5\%, 68.2\%, 68.4\%, and 82.3\% on the individual projects).
Other findings were classified as hints (3.0\%).

\vspace{0.03in}

%\paragraph*{\GROUMiner}
\GROUMiner is a misuse detector for Java~\cite{NNP+09}.
It creates a graph-based object-usage representation (\GROUM) for each target method.
A \GROUM is a directed acyclic graph whose nodes represent method calls, branchings, and loops and whose edges encode control and data flows.
\GROUMiner performs frequent-subgraph mining on sets of such graphs to detect recurring usage patterns with a minimum support of $6$.
% Detection
When at least 90\% of all occurrences of a sub-pattern can be extended to a larger pattern, but some cannot, those \emph{rare} inextensible occurrences are considered as violations.
Note that such violations have always exactly one node less than a pattern.
The detection of patterns and violations happens at the same time.
Violations are ranked by their \emph{rareness}, i.e., the support of the pattern over the support of the violation.
% Capabilities
\GROUMiner detects missing method calls.
It also detects missing conditions and loops at the granularity of a missing branching or loop node.
However, it cannot consider the actual condition.
% Performance
The evaluation applied \GROUMiner to nine projects individually, thereby finding violations of project-specific patterns.
The authors reviewed the top-10 violations per project, a total of 184 findings, and found $5.4\%$ true positives (three times 0\%, five times 6.7\%, and once 7.8\% on the individual projects).
Other findings were classified as code smells (7.6\%) or hints (6.0\%).

\vspace{0.03in}

%\paragraph*{DMMC}~\cite{MM13}
\DMMC is a misuse detector for Java~\cite{MBM10}, specialized in missing method calls.
The detection is based on type usages, i.e., sets of methods called on an instance of a given type in a given method.
% and the signature of the method the usage occurred in (i.e., the context)
Two usages are \emph{exactly similar} if their respective sets match and are \emph{almost similar} if one of them contains exactly one additional method.
% Detection
The detection is based on the assumption that violations should have only few exactly-similar usages, but many almost-similar ones.
The likelihood of a usage $x$ being a violation is expressed in the \emph{strangeness score} $= 1 - |E(x)| / (|E(x)| + |A(X)|)$, where $E(x)$ is the set of usages that are exactly similar to $x$ and $A(x)$ the set of those that are almost similar.
A usage is considered a violation if its strangeness score is above $0.97$.
Violations are ranked by the strangeness score.
% Capabilities
\DMMC detects misuses with exactly one missing method-call.
% Performance
The evaluation applied \DMMC to a single project, thereby finding project-specific violations.
The authors manually reviewed all findings with a strangeness score above 97\%, a total of 19 findings, and confirmed 73.7\% as true positives.
The evaluation was repeated later~\cite{MM13}, applying \DMMC to three projects individually, thereby finding project-specific violations for a predefined set of APIs.
The authors report that they manually reviewed approximately 30 findings, and confirmed 17 ($\approx 56.7\%$) as true positives.
Others were classified as workarounds for bugs inside a used API.

\vspace{0.03in}

%\paragraph*{\Tikanga}~\cite{WZ11}
\Tikanga is a misuse detector for Java~\cite{WZ11} that builds on
%the same algorithm as
\Jadet.
It extends the simple call-order properties
% of the form \code{m() $\prec$ n()} 
to general Computation Tree Logic
% (CTL) 
formulae on object usages.
Specifically, it uses formulae that require a certain call to occur,
% in a usage, 
formulae that require two calls in
%a certain
order, and formulae that require a certain call to happen after another.
It uses model checking to determine
% the subset of 
all those formulae with a minimum support of 20 in the codebase.
Violations are ranked by the \emph{conviction} measure~\cite{BMUT97} of the association between the set of present formulae and the set of missing formulae in the violating usage.
% Detection
It then applies Formal Concept Analysis~\cite{GW99} to obtain patterns and violations at the same time.
% Capabilities
\Tikanga's capabilities are the same as \Jadet's.
%, but it also detects violations of patterns with two or less calls.
% Performance
The evaluation applied \Tikanga to six projects individually, finding violations of project-specific patterns.
The authors manually reviewed the top-25\% of findings per project, a total of 121 findings, and confirmed 9.9\% as true positives (0\%, 0\%, 8.3\%, 20.0\%, 21.4\%, and 33.3\% on the individual projects).
Other findings were classified as code smells (29.8\%).

\vspace{0.03in}

%\paragraph*{\DroidAssist}~\cite{NPVN16}
\DroidAssist is a detector for Android Java Bytecode~\cite{NPVN16}.
It generates method-call sequences from source code and learns a Hidden Markov Model from them, %, using a modified version of the Baum-Welch algorithm~\cite{todo}.
to compute the likelihood of a particular call sequence.
If the likelihood is too small, the sequence is considered a violation.
\DroidAssist then explores different modifications of the sequence (adding, replacing, and removing calls) to find a slightly modified, more likely sequence.
% Capabilities
This allows it to detect missing and redundant method calls and even to suggest solutions for them.
An evaluation of this mechanism is not provided in the respective paper.

\vspace{0.04in}
\noindent{\textbf{Summary.}}
All detectors use code (snippets) as training and verification input.
Some require the code in a compiled format, such as Java Bytecode, while others directly work on source code.
Detectors typically encode usages as sets, sequences, or graphs.
Graph representations seem promising for simultaneously encoding usage elements, order, and data-flow relations.
With the exception of \DroidAssist and \DMMC, detectors mine patterns through frequent-itemset/subsequence/subgraph mining, according to their usage representation. 
To detect violations, they mine in-extensible parts of patterns that are themselves observed infrequently.
This implies that they cannot detect redundant elements, since a usage with such an element is never part of any pattern.
The exception is \DroidAssist, which might find redundant calls as being unlikely.

\autoref{tab:detector_capabilities} summarizes the detectors' capabilities with respect to \MUC.
Overall, we find that detectors cover only a small subset of all API-misuse categories.
While all detectors may, to some degree, identify missing method calls, only four detectors may identify missing \code{null} checks and missing value-or-state conditions, only three may identify missing iterations, and only two may identify missing exception handling.
None of the detectors targets all of these categories.

Existing detectors use both absolute and relative minimum support thresholds to identify patterns.
The exceptions are, again, \DroidAssist and \DMMC, which use probabilistic approaches.
Since many detectors produce a high number of false positives, they use a variety of ranking strategies.
Most of these rely mainly on the pattern support, but some use different concepts, such as \emph{rareness}, \emph{strangeness}, or \emph{conviction}.
A comparison of different ranking strategies is not reported in any of the publications.

\autoref{tab:survey-evaluations} summarizes the empirical evaluations of the surveyed detectors, as reported in their original papers.
Most evaluations apply detectors to target projects individually.
In this setting, the detectors learn project-specific patterns and identify respective violations
The number of projects ranges from \checkNum{1 to 20} (average 5.3; median 5).
The concrete projects samples are all distinct and mostly even disjunct.

To assess the detection performance, most authors review the top-X findings of their detectors, where X is a fixed number or percentage.
They then either present anecdotal evidence of true positives or measure the precision of detectors.
Many evaluations also present additional categories of findings, such as code smells, to distinguish false positives from other non-misuse findings that may still be valuable to developers.
The definitions of when a finding belongs to which category---if provided---differ between publications, even if they use the same label, e.g., ``bug'' or ``code smell.''
No evaluation considers the recall of the respective detector.

Overall, it appears that the detectors that focus on specific violations, such as error handling or missing method calls, have higher precision.
However, simply comparing detectors based on their reported empirical results would be unreliable, since the target projects, the review sample sizes, and the criteria to assess detector findings differ between the studies.

%% file: sections/table-capabilities.tex
%!TEX root = ../paper.tex

\begin{table}[tb]
  \newcommand*\FC{{\hspace{.5mm}\color{green!70!black}$\CIRCLE$\hspace{.5mm}}}
  \newcommand*\MC{{\hspace{.5mm}\color{green!70!black}$\LEFTcircle$\hspace{.5mm}}}
  \newcommand*\MU{{\hspace{.5mm}\color{red}$\RIGHTcircle$\hspace{.5mm}}}
  \newcommand*\FU{{\hspace{.5mm}\color{red}$\Circle$\hspace{.5mm}}}

  \centering
  \caption{Capabilities of Surveyed API-Misuse Detectors.
  \FC~denotes the capability to detect a violation.
  \MC~denotes the capability to detect a violation under special conditions.
  % \MU~denotes that a detector might detect a violation as a side effect of its other capabilities.
  \FU~denotes the inability to detect a violation.}
  \resizebox{\columnwidth}{!}{
  \begin{tabular}{@{}l*{11}c @{}}
    \toprule
       & \multicolumn{2}{c}{Method Calls} & \multicolumn{5}{c}{Conditions} & \multicolumn{2}{c}{Ex. Handl.} & \multicolumn{2}{c}{Iteration} \\
    \cmidrule{2-3} \cmidrule(lr){4-8} \cmidrule(r){9-10} \cmidrule(l){11-12}
      Misuses Detector &
        \rot{Missing} &
        \rot{Redundant} &
        \rot{Missing \code{null}} &
        \rot{Missing Val./State} &
        \rot{Missing Sync.} &
        \rot{Missing Context} &
        \rot{Redundant} &
        \rot{Missing} &
        \rot{Redundant} &
        \rot{Missing} &
        \rot{Redundant} \\
    \midrule
      \PRMiner~\cite{LZ05}       & \FC & \FU & \FU & \FU & \FU & \FU & \FU & \FU & \FU & \FU & \FU \\
      \Chronicler~\cite{RGJ07b}  & \FC & \FU & \FU & \FU & \FU & \FU & \FU & \FU & \FU & \FU & \FU \\
      \Colibri~\cite{L07}        & \FC & \FU & \FU & \FU & \FU & \FU & \FU & \FU & \FU & \FU & \FU \\
      \Jadet~\cite{WZL07}        & \FC & \FU & \FU & \FU & \FU & \FU & \FU & \FU & \FU & \MC & \FU \\
      \RGJ~\cite{RGJ07}          & \MC & \FU & \FC & \FC & \FU & \FU & \FU & \FU & \FU & \FU & \FU \\
      \Alattin~\cite{TX09b}      & \MC & \FU & \FC & \MC & \FU & \FU & \FU & \FU & \FU & \FU & \FU \\
      \AX~\cite{AX09}            & \MC & \FU & \MC & \MC & \FU & \FU & \FU & \FC & \FU & \FU & \FU \\
      \CARMiner~\cite{TX09}      & \MC & \FU & \FU & \FU & \FU & \FU & \FU & \FC & \FU & \FU & \FU \\
      \GROUMiner~\cite{NNP+09}   & \FC & \FU & \MC & \MC & \FU & \FU & \FU & \FU & \FU & \MC & \FU \\
      \DMMC~\cite{MBM10}         & \FC & \FU & \FU & \FU & \FU & \FU & \FU & \FU & \FU & \FU & \FU \\
      \Tikanga~\cite{WZ11}       & \FC & \FU & \FU & \FU & \FU & \FU & \FU & \FU & \FU & \MC & \FU \\
      \DroidAssist~\cite{NPVN16} & \FC & \FC & \FU & \FU & \FU & \FU & \FU & \FU & \FU & \FU & \FU \\
    \bottomrule
  \end{tabular}
  }
  \label{tab:detector_capabilities}
\end{table}

%% file: sections/table-survey-evaluation.tex
%!TEX root = ../paper.tex

\begin{table}[tb]
  \centering
  \caption{Summary of Empirical Evaluations of Surveyed API-Misuse Detectors.
  For the evaluation setup, IP denotes that detectors mine on the individual target projects and CP that they mine cross-project.}
  \resizebox{\columnwidth}{!}{
  \begin{tabular}{lrcl *{1}{r@{\hskip 0.05in}r}}
    \toprule
             & \# of Target                 & Eval. & \# of Reviewed & \\
    Detector & \multicolumn{1}{l}{Projects} & Setup &  Findings      & \multicolumn{2}{l}{Precision (Range)} \\
    \midrule
      \PRMiner~\cite{LZ05}       &     3 & IP & Top 60           & 18.1\% & (10-27\%)  \\ % 26.7\%, 10.0\%, 14.3\%
      \Chronicler~\cite{RGJ07b}  &     5 & IP & \multicolumn{3}{l}{{\color{gray}example-based}} \\
      \Colibri~\cite{L07}        &     5 & IP & \multicolumn{3}{l}{{\color{gray}example-based}} \\
      \Jadet~\cite{WZL07}        &     5 & IP & Top 10/project   &  6.5\% & (0-13\%)   \\ % 10.5\%, 13.3\%, 7.7\%, 0, 0
      \Jadet~\cite{GWZ10}        &    20 & CP & Top 25\% (50)    &  8.0\% & (0-100\%)  \\ % 100\%, 0, 4.5\%, 40\%, 0\%
      \RGJ~\cite{RGJ07}          &     1 & IP & \multicolumn{3}{l}{{\color{gray}example-based}} \\
      \Alattin~\cite{TX09b}      &     6 & CP & Top 10/project   & 29.5\% & (13-100\%) \\ % 26.2\%, 32.7\%, 28.1\%, 12.5\%, 100\%, 52.6\%
      \AX~\cite{AX09}            &     3 & IP & All (292)        & 90.4\% & (50-94\%)  \\ % 93.5\%, 50.0\%, 90.3\%
      \CARMiner~\cite{TX09}      &     5 & CP & Top 10/project   & 60.1\% & (41-82\%)  \\ % 68.4\%, 82.3\%, 54.5\%, 68.2\%, 41.1\%
      \GROUMiner~\cite{NNP+09}   &     9 & IP & Top 10/project   &  5.4\% & (0-8\%)    \\ % 3x 0, 5x 6.7\%, 7.8\%
      \DMMC~\cite{MBM10}         &     1 & IP & All (19)         & 73.7\% &            \\
      \DMMC~\cite{MM13}          &     3 & IP & Top 30           & 56.7\% &            \\
      \Tikanga~\cite{WZ11}       &     6 & IP & Top 25\% (121)   &  9.9\% & (0-33\%)   \\ % 33.3\%, 0, 8.3\%, 21.4\%, 20.0\%, 0
      \DroidAssist~\cite{NPVN16} & \multicolumn{4}{c}{{\color{gray}not evaluated}} \\
    \bottomrule
  \end{tabular}
  }
  \label{tab:survey-evaluations}
\end{table}

%% file: sections/benchmark.tex
%!TEX root = ../paper.tex

\input{sections/table-run-stats}

\section{Experimental Setup}
\label{sec:benchmark}

%As we see in \autoref{tab:survey_evaluation}, the common practice for evaluating an API-misuse detector is to apply it to a small number of real-world projects and to manually review the top-ranked findings.
%The reviewers then classify findings into true positives, false positives, and a varying set of further categories, such as code smells or hints for improvement.
%This procedure has two inherent problems: (1) Since the classification criteria vary and are not always clearly documented, reproduction of the results is not straightforward and often even infeasible, and (2) Since the detectors are applied to different sets of projects, we cannot compare them fairly to one another.

In \autoref{sec:misuse-detectors-related-work}, we conceptually compared detectors' capabilities.
In this section, we describe the experimental setup we use to \textit{empirically} compare their capabilities.
We design three experiments, to measure both the detectors' precision and recall.
We build these experiments on \MUBench as a ground-truth dataset.
This enables us to compare all detectors on the same target projects and with respect to the same known misuses.

\InlineSec{Subject Detectors}
In this study, we focus on misuse detectors for Java APIs, because \MUBench contains examples of Java-API misuses.
Our survey identifies \checkNum{seven such detectors}.
We contacted the respective authors and got responses from all of them.
However, we learned that we cannot run \CARMiner{} and \Alattin{}, because they both depend on Google Code Search, a service that is no longer available~\cite{farewellGoogle}.
We exclude \DroidAssist{}, because its implementation only supports Dalvik Bytecode,\footnote{A bytecode format developed by Google, which is optimized for the characteristics of mobile operating systems (especially for the Android~platform).} while the examples in \MUBench are general Java projects, which compile to Java Bytecode.
%while our benchmark contains many other Java projects. 
This leaves us with four detectors \Jadet{}, \GROUMiner{}, \Tikanga{}, and \DMMC{}.

\InlineSec{Misuse Dataset}
We use \MUBench, described in~\autoref{sec:definitions}, to find targets for our evaluations.
While \GROUMiner{} works on source code, \Jadet{}, \Tikanga{}, and \DMMC{} require Java Bytecode as input.
Thus, we can only compare them on project versions for which we have both source code and Bytecode.
Since Bytecode is not readily available for most project versions in the dataset, we resort to compiling them ourselves by adding necessary build files and fixing any dependency issues.
We exclude \checkNum{26 project versions (47\%)} with compilation errors that we could not fix.
In the end, we have \checkNum{29 compilable project versions} and \checkNum{25 hand-crafted examples}, with \checkNum{64 misuses} in total, for our experiments.
Note that some project versions contain multiple misuses.
The last three rows in \autoref{tab:datasets} describe the subsets of this dataset that we use in the individual experiments.
We publish the dataset~\cite{mubench} for others to use in future studies.

\subsection{Experiment~P}
\label{e2}

We design \nameref{e2} to assess the precision of detectors.

\InlineSec{Motivation}
Past studies show that developers rarely use analysis tools that produce many false positives~\cite{FLLNSSLNSS02,BBCCFHHKME10,JS13}.
Therefore, for a detector to be adopted in practice, it needs a high precision.

\InlineSec{Setup}
To measure precision, we follow the most-common experimental setup we found in the literature (cf. \autoref{tab:survey-evaluations}).
First, we run detectors on individual project versions.
In this setting, they mine patterns and detect violations on a per-project basis.
Second, we manually validate the \checkNum{top-20} findings per detector on each version, as determined by the respective detector's ranking strategies.
We limit the number of findings, because it seems likely that developers would only consider a fixed number of findings, rather than all of a potentially very large number of findings.
Hence, the precision in a detector's top findings is likely crucial for tool adoption.
Also, we need to limit the effort of reviewing findings of multiple detectors on each project version.

\InlineSec{Dataset}
Since manually reviewing findings of all detectors on all project versions is infeasible, we sample \checkNum{five project versions}.
To ensure a fair selection of projects, we first run all detectors on all project versions.
For practical reasons, we timeout each detector on an individual project version after \checkNum{two hours}.
The run statistics are summarized in \autoref{tab:run-stats}.

\input{sections/table-run-findings-correlation}

\Jadet and \Tikanga fail on \checkNum{one project version} and \DMMC fails on \checkNum{four project versions}, since the Bytecode contains constructs that the detectors' respective Bytecode toolkits do not support.
\GROUMiner times out on \checkNum{eight project versions} and produces an error on \checkNum{one other version}.
We exclude any project version where a detector fails.

For the remaining \checkNum{15 versions}, we observe that the total number of findings correlates across detectors.
\autoref{tab:run-findings-correlation} shows that the pairwise correlation (Pearson's $r$) is strong ($\geq 0.75$) or medium ($\geq 0.5$) for all pairs of detectors, except for \Jadet and \GROUMiner ($r = 0.49$).
This means that either all detectors report a relatively large or a relatively small number of findings on any given project version.
We hypothesise that the total number of findings might be related to the detectors' ability to precisely identify misuses in a given project version.
Therefore, we sample project versions according to the average normalized number of findings across all detector.
We normalize the number of findings per detector on all project versions by the maximum number of findings of that detector on any project version.
We sample the two projects with the highest average normalized number of findings across all detectors (\aName{Closure}~\cite{closure} v319 and \aName{iText}~\cite{itext} v5091) and the two projects with the lowest average normalized number of findings across all detectors (\aName{JMRTD}~\cite{jmrtd} v51 and \aName{Joda-Time}~\cite{jodatime} v1231).
Additionally, we randomly select one more project version (\aName{Apache Lucene}~\cite{lucene} v1918) from the remaining projects, to cover the middle ground.
Note that we select at most one version from each distinct project, because different versions of the same project may share a lot of code, such that detectors are likely to perform similarly on them.
This dataset for \nameref{e2} is summarized in Row \checkNum{3} of \autoref{tab:datasets}.

\InlineSec{Metrics}
We calculate the precision of the detector, i.e., the ratio between the number of true positives over the number of findings.

\InlineSec{Review Process}
Two authors independently review each of the \checkNum{top-20} findings of the sampled project versions and mark it as a misuse or not.
To determine this, they consider the logic and the documentation in the source code, the API's documentation, and its implementation if publicly available.
After the review, any disagreements between the reviewers are discussed until a consensus is reached.
We report Cohen's Kappa score as a measure of the reviewers' agreement.
Note that we follow a lenient reviewing process.
For example, assume a usage misses a check \code{if (iterator.hasNext())} before calling \code{iterator.next()}.
If the detector finds that \code{hasNext()} is missing, we mark the finding as a hit, even though this does not explicitly state that the call to \code{next()} should be guarded by a check on the return value of \code{hasNext()}.
This follows our intuition that such findings may still provide a developer with a valuable hint about the problem.

\subsection{Experiment~RUB}
\label{e1}

We design \nameref{e1} to assess the detection capabilities of our subject detectors, i.e., to measure an upper bound to their recall under the assumption that they always mine the required pattern.

\InlineSec{Motivation}
We argue that it is important for developers to know which misuses a particular tool may or may not find, in order to decide whether the tool is adequate for their use case and whether they must take additional measures.
Moreover, it is important for researchers to know which types of misuses existing detectors may identify, in order to direct future work.
Therefore, we measure detectors' recall while providing sufficiently many correct usages that would allow them to mine the required pattern.

\InlineSec{Dataset}
For this experiment, we use all compilable project versions from the \MUBench dataset with the respective known misuses, as well as the hand-crafted misuse examples.
This dataset for \nameref{e1} is summarized in Row \checkNum{4} of \autoref{tab:datasets}.

\InlineSec{Setup}
Recall that all our subject detectors mine patterns, i.e., frequently reoccurring API usages, and assume that these correspond to correct usages.
They use these patterns to identify misuses.
Recall further that each detector has a distinct representation of usages and patterns and its own mining and detection strategies.
If a detector fails to identify a particular misuse, this may be due to
\begin{enumerate*}[label=(\arabic*)]
  \item an inherent limitation of the detector, e.g., because it cannot represent some usage element such as conditions, or
  \item a lack of examples of respective correct usage for pattern mining, i.e., a limitation of the training data.
\end{enumerate*}
With \nameref{e1}, we focus on (1), i.e., we take (2) out of the equation and assess the detectors' general ability to identify misuses.
To this end, we provide the detectors with sufficiently many examples of correct usage corresponding to the misuses in question.
This guarantees that they could mine a respective pattern.
If the detector is unable to identify a misuse in this setting, we know the problem lies with the detector itself.

We manually create a correct usage for each misuse in the dataset, using the fixing commits recorded in \MUBench.
For each misuse, we take the entire code of the method with the misuse after the fixing commit and remove all code that has no data or control dependencies to the objects involved in the misuse.
We store the code of this \emph{crafted correct usage} in our dataset.

In the experiment, we run each detector once for each individual known misuse in the dataset.
In each run, we provide the detector with the file that contains the known misuse and with \checkNum{50} copies of the respective crafted correct usage.
We ensure that the detector considers each copy as a distinct usage.
We configure the detectors to mine patterns with a minimum support of \checkNum{50}, thereby ensuring that they mine patterns only from the code in the crafted correct usage.
We chose \checkNum{50} as a threshold, since it is high enough to ensure that no detector mines patterns from the code in the file with the misuse.

\InlineSec{Metrics}
We calculate two numbers for each detector.
The first is its  \emph{conceptual recall upper bound}, which is the fraction of the known misuses in the dataset that match its capabilities from \autoref{tab:detector_capabilities}.
Note that the conceptual recall upper bound is calculated offline, without running any experiments.
The second is the detector's \emph{empirical recall upper bound}, which is the fraction of misuses a detector actually finds from all the known misuses in the dataset.
An ideal detector should have an empirical recall upper bound equal to its conceptual recall upper bound.
Otherwise, its practical capabilities do not match its conceptual capabilities.
In such cases, we investigate the root causes for such mismatches.
Note that we use the term ``upper bound,'' because neither recall rate reflects the detectors' recall in a setting without guarantees on the number of correct usages for mining.

\InlineSec{Review Process}
To evaluate the results, we review all \emph{potential hits}, i.e., findings from each detector that identify violations in the same files and methods as known misuses.
Two authors independently review each such potential hit to determine whether it actually identifies one of the known misuses.
If at least one potential hit identifies a misuse, we count it as a \emph{hit}.
After the review, any disagreements between the reviewers are discussed until a consensus is reached.
We report Cohen's Kappa score as a measure of the reviewers' agreement.
We follow the same lenient review process as for \nameref{e2}.

\subsection{Experiment~R}
\label{e3}

We design \nameref{e3} to assess the recall of detectors.

\InlineSec{Motivation}
While \nameref{e1} gives us an upper bound to the recall of misuse detectors, we also want to assess their actual recall where we do not provide them with correct usages ourselves.
Due to the lack of a ground-truth dataset, such an experiment has not been attempted before in any of the misuse-detection papers we surveyed.

\InlineSec{Dataset}
As the ground truth for this experiment, we use all known misuses from real-world projects in \MUBench plus the true positives identified by any of the detectors in \nameref{e2}.
This means that~\nameref{e3} not only evaluates recall against the misuses of \MUBench, but also practically cross-validates the detector capabilities against each other.
We exclude the hand-crafted misuse examples from this experiment, since there is no corresponding code for the detectors to mine patterns from.
The dataset we use for \nameref{e3} is summarized in Row \checkNum{5} of~\autoref{tab:datasets}.

\InlineSec{Setup}
We run all detectors on all projects versions individually, i.e., we use the same per-project setup as for \nameref{e2}.

\InlineSec{Metrics}
We calculate the recall of the detectors, i.e., the number of actual hits over the number of known misuses in the dataset.

\InlineSec{Review Process}
We review all potential hits in the same process as for \nameref{e1}.
This gives us the detectors' recall with respect to a large number of known misuses from \MUBench.

%% file: sections/table-run-stats.tex
%!TEX root = ../paper.tex

\begin{table*}[tb]
  \newcommand*\ERR{{\color{red!70!white}error}}
  \newcommand*\TO{{\color{red!70!white}timeout (2h)}}
  \definecolor{lightblue}{rgb}{0.68, 0.85, 0.9}
  
  \centering
  \small
  \caption{Number of Findings per Detector on All Compilable Project Versions in \MUBench.
  \nameref{e2} includes the two projects with the highest number of findings, the two projects with the lowest number of findings, and one randomly selected project.}
  \begin{tabular}{lrrrrrrl}
    \toprule
            &                  & \multicolumn{5}{c}{Number of Findings} & \multirow{2}{*}{\begin{tabular}[b!]{@{}l@{}}Sample\\Criterion\end{tabular}} \\
                                 \cmidrule(l){3-7}
    Project & Version & \Jadet & \GROUMiner & \Tikanga & \DMMC & Norm. Avg. \\
    \midrule
    \aName{Apache Commons Lang} &    587 &     0 &    28 &    0 &   157 & 0.06 & \\
    \aName{Apache Commons Math} &    998 &    17 &  \ERR &   17 &   686 & 0.20 & \\
    \aName{ADempiere}           &   1312 &     0 &    27 &    0 &   116 & 0.05 & \\
    \aName{Alibaba Duid}        & e10f28 &    17 &   \TO &    5 &   520 & 0.13 & \\
    \aName{Closure}             &    114 &   113 &   101 &   24 &  1233 & 0.49 & \\
    \rowcolor{orange!30!white}
    \aName{Closure}             &    319 &   176 &   126 &   45 &  1945 & 0.74 & highest \\
    \aName{Closure}             &    884 &    71 &   167 &   33 &  1966 & 0.63 & \\
    \aName{Apache HttpClient}   &    302 &     0 &    12 &    0 &   114 & 0.03 & \\
    \aName{Apache HttpClient}   &    444 &     0 &    15 &    0 &   110 & 0.03 & \\
    \aName{Apache HttpClient}   &    452 &     0 &    12 &    0 &   113 & 0.03 & \\
    \rowcolor{orange!30!white}
    \aName{iText}               &   5091 &    17 &   198 &   55 &  1138 & 0.55 & highest \\
    \aName{Apache Jackrabbit}   &   1601 &    12 &   186 &   22 &  \ERR & 0.41 & \\
    \aName{Apache Jackrabbit}   &   1678 &     0 &    15 &    0 &  \ERR & 0.03 & \\
    \aName{Apache Jackrabbit}   &   1694 &    13 &   186 &   22 &  \ERR & 0.41 & \\
    \aName{Apache Jackrabbit}   &   1750 &    10 &   \TO &    8 &   434 & 0.12 & \\
    \aName{JFreeChart}          &    103 &   167 &   \TO &   88 &   673 & 0.69 & \\
    \aName{JFreeChart}          &    164 &   168 &   \TO &   90 &   664 & 0.69 & \\
    \aName{JFreeChart}          &    881 &   194 &   \TO &   93 &   745 & 0.76 & \\
    \aName{JFreeChart}          &   1025 &   194 &   \TO &   93 &   747 & 0.76 & \\
    \aName{JFreeChart}          &   2183 &   190 &   \TO &  100 &   906 & 0.81 & \\
    \aName{JFreeChart}          &   2266 &   195 &   \TO &  102 &   913 & 0.82 & \\
    \rowcolor{lightblue!40!white}
    \aName{JMRTD}               &     51 &     0 &    11 &    0 &    29 & 0.02 & lowest \\
    \aName{JMRTD}               &     67 &     0 &    10 &    0 &    35 & 0.02 & \\
    \rowcolor{lightblue!40!white}
    \aName{Joda-Time}           &   1231 &     0 &     0 &    0 &     1 & 0.00 & lowest \\
    \aName{Apache Lucene}       &    207 &     0 &   140 &    0 &   182 & 0.20 & \\
    \aName{Apache Lucene}       &    754 &     0 &    54 &    0 &   265 & 0.10 & \\
    \aName{Apache Lucene}       &   1251 &     2 &    62 &    0 &  \ERR & 0.11 & \\
    \rowcolor{green!30!white}
    \aName{Apache Lucene}       &   1918 &     2 &    88 &    4 &   583 & 0.20 & random \\
    \aName{Mozilla Rhino}       & 286251 &  \ERR &    55 & \ERR &   257 & 0.20 & \\
    \bottomrule
  \end{tabular}
  \label{tab:run-stats}
\end{table*}

%% file: sections/table-run-findings-correlation.tex
%!TEX root = ../paper.tex

\begin{table}[tb]
  \newcommand\strong[1]{\textbf{#1}}
  \newcommand\medium[1]{\emph{#1}}
  \newcommand\weak[1]{#1}
  
  \centering
  \caption{Correction of the Number of Findings per Project Version For All Pairs of Detectors (Pearson's $r$).
  Strong correlation ($r \geq 0.75$) in \strong{bold}.
  Medium correlation ($r \geq 0.5$) in \medium{italic}.}
  \resizebox{\columnwidth}{!}{
  \begin{tabular}{lrrrr}
  \toprule
    & \Jadet & \GROUMiner & \DMMC & \Tikanga \\
  \midrule
  % considering only the 15 project version where all detectors succeed
  \Jadet     & \strong{1.00} \\
  \GROUMiner &   \weak{0.49} & \strong{1.00} \\
  \DMMC      & \strong{0.85} & \strong{0.78} & \strong{1.00} \\
  \Tikanga   & \medium{0.70} & \strong{0.82} & \strong{0.88} & \strong{1.00} \\
  % considering all project version where the two compared detectors succeed
  %\Jadet     & 1.00	\\
  %\GROUMiner & 0.38 &	1.00 \\
  %\DMMC      & 0.59 & 0.78 & 1.00 \\
  %\Tikanga   & 0.93 & 0.79 & 0.50 & 1.00 \\
  \bottomrule
  \end{tabular}
  }
  \label{tab:run-findings-correlation}
\end{table}

%% file: sections/pipeline.tex
%!TEX root = ../paper.tex

\section{\MUPipe}
\label{sub:the_pipeline}

To systematically assess and compare API-misuse detectors, we built \MUPipe, a benchmarking pipeline for API-misuse detectors.
\MUPipe automates large parts of the experimental setup presented in \autoref{sec:benchmark} and facilitates the reproduction of our study.
It also enables adding new detectors to the comparison, as well as benchmarking with different or extended datasets, in the future.
We publish the pipeline~\cite{mubench} for future studies.

\subsection{Automation}

Following the idea of automated bug-detection benchmarks for C programs, such as~\aName{BugBench}~\cite{LLQ+05} and \aName{BegBunch}~\cite{CHKL+09}, we facilitate the benchmarking of multiple detectors on our misuse dataset with an evaluation pipeline.
\MUPipe automates many of our evaluation steps, such as retrieval and compilation of target projects, running detectors, and collecting their findings.
\MUPipe provides a command-line interface to control these steps.
We subsequently describe the pipeline steps we implemented to facilitate our evaluation.

\InlineSec{Checkout}
\MUPipe uses the recorded commit Id from \MUBench to obtain the source code of the respective project version.
It supports SVN and Git repositories, source archives (zip), as well as a special handling for the hand-crafted examples that come with \MUBench.

\InlineSec{Compile}
For every project version, \MUPipe first copies the entire project source code, the individual files containing known misuses, and the respective crafted correct usages for \nameref{e1} each into a separate folder.
It then uses the respective build configuration from the dataset to compile all Java sources to Bytecode.
After compilation, it copies the entire project Bytecode, the Bytecode of the individual files containing known misuses, and the Bytecode of the respective crafted correct usages each into a separate folder.
This way, we may provide the detectors with the source code or Bytecode of each of these parts individually.

\InlineSec{Detect}
For each detector, we also built a \emph{runner} to have a unified command-line interface for all detectors.
For every project version, \MUPipe invokes the detector with the paths to the respective source code and Bytecode.
%The runners use these to provide the right input to the respective detector and to output its findings. 
All detectors are invoked with the best configuration reported in their respective publication.
Apart from adding some accessor methods that allow us to obtain the detectors' output, all detector implementations were left unchanged.

%After the detection, the runners convert the detectors' output into a unified format for findings to facilitate the following validation step.
%For each finding, this format specifies the name of the file and the name of the method that the finding is in.
%In addition, the runners add tool-specific data that helps with validation (e.g., the detector's confidence value).

\InlineSec{Validation}
To help with the manual review of findings, \MUPipe automatically publishes experiment results to a review website~\cite{artifact-page}.
For every detector finding, the website shows the source code it is found in along with any metadata the detector provides, such as the violated pattern, the properties of the violation, and the detector's confidence.

For Experiments RUB and R, \MUPipe automatically filters potential hits, by matching findings to known misuses by file and method name.
On the review website, a reviewer sees the description of the known misuse as well as its fix, along with the set of potential hits that need to be reviewed.
For \nameref{e2}, \MUPipe shows all findings of the detector on the review site.

The review website allows reviewers to save an assessment and comment for each finding.
It also ensures at least two reviews for each finding, before automatically computing the experiment statistics, such as precision, recall, and Cohen's Kappa scores.

\subsection{Reproduction, Replication, and Extension}

\MUPipe comes with a Docker image, which allows running reproducible experiments across platforms, without the need to ensure a proper environment setup.
Its review website comes with a second Docker image, which allows serving it standalone.
Moreover, it is based on PHP and MySQL, such that it can be hosted on any off-the-shelf webspace.
The review website facilitates independent reviews, even when researchers work from different locations, while ensuring review integrity using authentication.
The website may also directly be used as an artifact to publish review results and experiment statistics.
\MUPipe defines a simple data schema for misuse examples to facilitate extensions of \MUBench.
It also provides a convenient Java interface as a Maven dependency to enable plugging in additional detectors for evaluation on the benchmark.
For further details on how~to~use~or extend \MUPipe, we refer the readers to our project website~\cite{mubench}.

%% file: sections/results.tex
%!TEX root = ../paper.tex

\section{Results}
\label{sec:results}

We now discuss the results of comparing \Jadet, \GROUMiner, \Tikanga, and \DMMC in our experiments.
All reviewing data is available on our artifact page~\cite{artifact-page}.

\subsection{\nameref{e2}}

\begin{table*}[tb]
  \centering
  \small
  \caption{\nameref{e2}: Precision of the Detectors on the \checkNum{Top-20 Findings} on \checkNum{5 Projects} and Root Causes for False Positives.}
  \input{sections/table-ex2-results}
  \label{tab:ex2-results}
\end{table*}

\autoref{tab:ex2-results} shows our precision results, based on reviewing the \checkNum{top-20 findings} per detector on each of our \checkNum{five sample projects}.
The second column shows the total number of reviewed findings, \checkNum{230} in total across all detectors.
Note that all detectors report less than \checkNum{20 findings} for some projects.
The third column shows the confirmed misuses after resolving disagreements, and the fourth column shows the precision with respect to the reviewed findings.
The fifth column shows the Kappa score for the manual reviews, and the remaining columns show the frequencies of root causes for false positives.
We find that the precision of all detectors is extremely low.
\Tikanga shows the best precision of only \checkNum{11.4\%}.
\Jadet and \DMMC follow immediately behind, with a precision of \checkNum{10.3\%} and \checkNum{9.9\%}, respectively.
\GROUMiner reports only false positives in its \checkNum{top-20 findings}.

\begin{obs}{e2-precision}
  All detectors have extremely low precision (below \checkNum{12\%}).
  On average, they report less than 1.5 actual misuses in their top-20 findings.
  %~\sn{where did you get this number from?}~\sa{the number was wrong. Its $17 \text{ confirmed misuses } / 230 \text{ reviewed findings } * \text{(top-)}20 = 1.478$}~\sn{So it is a precision of 1.5\% or the absolute number of true positives.. right now the number is confusing that it is the percentage but phrased as an absolute/average number of true positives}~\sa{I rephrased it. Better now?}
\end{obs}

The Kappa scores indicate high reviewer agreement, which shows that all detectors produced mostly clear false positives.
The score is a little lower for \Tikanga, because it reported one confirmed misuse twice, which one of the reviewers first accepted as an actual hit while the other did not.
The score is also lower for \DMMC, because we initially disagreed on several violations it identifies in \code{Iterator} usages that do not check \code{hasNext()}, but the underlying collection's size.

\subsubsection*{True Positives}

Out of the \checkNum{230 reported findings} we reviewed, we confirm \checkNum{17 true misuses}.
\DMMC reports \checkNum{8 misuses} of an iterator API where \code{hasNext()} is not checked.
\Jadet reports \checkNum{4 misuses} that access a collection without checking its size before.
Also for collections, \Tikanga reports \checkNum{4 misuses} with a missing \code{hasNext()} and \checkNum{1 misuse} with a missing size check.
One misuse is reported by both \Tikanga and \Jadet and another by both \Tikanga and \DMMC.
Additionally, \Jadet reports one misuse twice.
This leaves a total of \checkNum{14 unique misuses}, all different from the known misuses in \MUBench.
% \mm{Are you saying that you didn't find any of the misusages in MuBench? If this was true, than you come you found a recall of higher than 0? I am confused (which also might be due to being tired ...)}
%
Interestingly, \emph{all} these misuses are missing value or state conditions, for which the detectors report only missing calls to methods that should be used in the respective missing checks.
We accept these findings in our lenient review process.

\begin{obs}{e2-misuses-allCSV}
  All \checkNum{14 confirmed misuses} in \nameref{e2} are missing value or state condition checks before accessing the elements of a collection, either directly or through an iterator.
\end{obs}

\subsubsection*{False Positives}

To identify opportunities to improve the precision of misuse detectors, we systematically investigate the root causes for the false positives they report.
In the following, we discuss these root causes summarized across all detectors, in the order of their absolute frequency.

\vspace{0.03in}
\noindent {\bf 1. Uncommon.}
% description
Particular usages may violate the patterns that detectors learn from frequent usages, without violating actual API usage constraints.
Detectors cannot differentiate infrequent from invalid usage.
% example
For example, \DMMC and \Jadet learn that the methods \code{getKey()} and \code{getValue()} of \code{MapEntry} usually appear together in code.
They both report violations if a call to either of these methods is missing, or, in case of \Jadet, if the calls appear in a different order.
However, there is no requirement by the API to always call both getter methods, let alone in a specific order.
Across the reported violations we analyzed, the detectors falsely report \checkNum{42 missing method calls} in cases where one out of a number of getter methods is missing or invoked in a different order.
Another example is that \Jadet and \Tikanga learn that methods such as \code{List.add()} and \code{Map.put()} are usually invoked in loops and report \checkNum{five missing iterations} for respective invocations outside a loop, which are perfectly fine according to the API.
Approaches such as multi-level patterns~\cite{SBA+15} or \Alattin's alternative patterns~\cite{TX09b} may help to mitigate this problem.
Also note that the four detectors in our experiments all use absolute frequency thresholds, while some of the detectors from our survey in \autoref{sec:misuse-detectors-related-work} also used relative thresholds.
Future work should investigate how these two alternatives compare.

\begin{obs}{e2-frequent}
  Particular usages may be uncommon without violating API constraints.
  Neglecting this causes \checkNum{73 (34.3\%)} of the~detectors' false positives in their \checkNum{top-20 findings}.
This calls for research on detecting patterns without setting a hard threshold on occurrence frequencies.
Meanwhile, relaxing requirements on the co-occurrence of getter methods might reduce false positives significantly.
\end{obs}

\vspace{0.03in}
\noindent {\bf 2. Analysis.}
% description
The detectors use static analysis to determine the facts that belong to a particular usage.
Imprecisions of these analyses lead to false positives.
% example
For example, the detectors mistakenly report \checkNum{five missing elements} in code that uses multiple aliases for the same object and another \checkNum{17} in code with nested control statements.
In both cases, the analysis failed to capture all calls belonging to the same usage.
\GROUMiner reports \checkNum{two missing method calls}, because it cannot resolve the receiver types in the chained calls and, therefore, fails to match a call between the pattern and the usage.
% example: interpocedural
Another example is that the detectors report \checkNum{eight missing method calls} due to chained calls on a fluent API, such as \code{StringBuilder}, where their analyses cannot determine that all calls actually happen on the same object.
\Jadet, \GROUMiner, and \DMMC together report \checkNum{nine missing calls} that happen transitively in a helper method of the same class or through a wrapper object, such as a \code{BufferedStream}.
\DMMC reports \checkNum{a missing call} that is located in the enclosing method of an anonymous class instance and \checkNum{a missing \code{close()} call} on a parameter that is, by contract, closed by the callers.
Moreover, \GROUMiner reports \checkNum{four missing conditions} that are checked by assertion helper methods.
An inter-procedural detection strategy, as proposed by \PRMiner~\cite{LZ05}, could mitigate this problem.
% one additional case is where the same object is retrieved twice via get, so same as chained calls on fluent API

\begin{obs}{e2-analysis}
  Imprecisions of the detectors' static analyses cause \checkNum{51 (23.9\%)} of the false positives in their \checkNum{top-20 findings}.
  An inter-procedural detection strategy might be able to eliminate \checkNum{14 (6.6\%)} of these false positives.
\end{obs}

\vspace{0.03in}
\noindent {\bf 3. Alternative.}
%~\mm{Is this really different from "Frequent"? Aren't the alternative correct usages non-frequent usages?}~\sa{They don't necessarily are, there might be multiple frequent alterantives and they would still trigger this problem. However, we could double check whether we have such a case at all. I'm not sure.}
% description
The detectors often learn a pattern and then report instances of alternative usages as violations.
We define \emph{alternative usages} as a different functionally correct way to use an API, either to achieve the same or a different functionality.
Note that multiple alternatives may occur frequently enough to induce patterns.
% example
For example, \Jadet, \Tikanga, and \DMMC learn that before a call to \code{next()}, there should always be a call to \code{hasNext()} on an \code{Iterator}.
Consequently, they report \checkNum{16 violations} in usages that check either \code{isEmpty()} or \code{size()} on the underlying collection before fetching only the first element through the \code{Iterator}.
\DMMC reports \checkNum{another violation}, because \code{isEmpty()} is used instead of \code{size()} before accessing a \code{List}.
Another example is that \Jadet, \Tikanga, and \DMMC learn that collections are filled one element at a time, e.g., by calling \code{add()}, and report \checkNum{10 missing methods} in usages that populate a collection differently, e.g., through the constructor or using \code{addAll()}.
%
% Another example is that \Jadet and \Tikanga learn that the size of a collection is usually retrieved in a loop header, when iterating over the list.
% Consequently, they each report one violation where the size is assigned to an intermediate variable before the loop and checked in the loop header, even though this usage is equivalent---assuming that the collection is not modified in the loop.
%
\GROUMiner reports \checkNum{four usages} where an alternative control statement is used, e.g., a \code{for} instead of a \code{while}.

%\paragraph*{Obtain}
% description
A special case of this root cause is alternatives to obtain an instance of a type.
% example
For example, \GROUMiner mistakenly reports \checkNum{two missing constructor calls} where the instance is not created through a constructor call as in the pattern, but returned from a method call.
\Jadet and \DMMC{} \checkNum{each report one missing constructor call} where an instance is not created, but passed as a parameter.
While handling alternative patterns is an open problem, some tools such as \Alattin already propose possible~solutions~\cite{TX09b}.

\begin{obs}{e2-alternative}
  A violation of a pattern might be an instance of an alternative, correct way to use the respective API.
  Not considering this causes \checkNum{41 (19.2\%)} of the false positives in their \checkNum{top-20 findings}.
\end{obs}

\vspace{0.03in}
\noindent {\bf 4. Inside.}
% description
Objects that are stored in fields are often used across multiple methods of the field's declaring class.
The respective API usages inside the individual methods might then deviate from usage patterns without being actual misuses.
% example
\autoref{lst:partial-usage} shows an example of such a case, where two fields of type \code{Iterator}, \code{in} and \code{out}, are used to implement the class \code{NeighborIterator}.
When \code{in} yields no more elements (Line~\ref{line:in-empty}), the call to \code{next()} in Line~\ref{line:unsafe-next} happens on \code{out} without a prior check whether it has more elements.
While this appears to be a misuse of the \code{Iterator} API inside the enclosing method, it is a correct usage inside the enclosing class, since \code{NeighborIterator} itself implements \code{Iterator} and, thereby, inherits its usage constraints.
Correct usages of \code{NeighborIterator} need to check its \code{hasNext()} method (Line~\ref{line:hasNext}) before calling its \code{next()} method (Line~\ref{line:next}), which ensures that \code{out} has more elements when \code{next()} is called on it.
\DMMC and \GROUMiner report \checkNum{sixteen violations} for such usages of fields of a class.
% Thereof, \DMMC reports \checkNum{six violations} in constructors where fields are initialized, but the respective objects not used otherwise.
% Similarly, \DMMC and \GROUMiner report \checkNum{three violations} in named constructors, i.e., methods that only create, initialize, and return an object, but not use it otherwise.
%
% Finally, \GROUMiner reports \checkNum{three violations} where it mines the patterns from usages in other methods of the same class and finds a deviance in a certain method, e.g., while most methods of \code{RAMDirectory} check whether the underlying file is open and throw an exception if it is not, the method \code{fileExists()} deviates from this pattern in that it does not throw an exception, but rather returns a boolean.

A special case of this root cause is when a class uses part of its own API in its implementation.
For example, when a \code{Collection} calls its own \code{add()} method in the implementation of its \code{addAll()} method.
\DMMC and \GROUMiner report \checkNum{four such violations}.
This is particularly interesting, because these are actually self usages of the API, while the detectors target client usages.
Since any codebase likely contains such self usages, detectors should consider this.

\begin{figure}[tb]
  \begin{lstlisting}[language=java,breaklines=true,escapechar=/]
class NeighborIterator implements Iterator<GraphNode> {
  private final Iterator<DiGraphEdge> in = ...;
  private final Iterator<DiGraphEdge> out = ...;
  
  @Override
  public boolean  hasNext() {/\label{line:hasNext}/
    return in.hasNext() || out.hasNext();
  }

  @Override
  public GraphNode next() {/\label{line:next}/
    boolean isOut = !in.hasNext();/\label{line:in-empty}/
    Iterator<DiGraphEdge> curIterator =  isOut ? out : in;
    DiGraphEdge s = curIterator.next();/\label{line:unsafe-next}/
    return isOut ? s.getDestination() : s.getSource();
  }
  
  ...
}
  \end{lstlisting}
  \caption{Correct Usages of \code{Iterator} Instances in the \aName{Closure} Project that Violate Usage Patterns.}
  \label{lst:partial-usage}
\end{figure}

\begin{obs}{e2-inside}
  The implementation code of a class may contain partial usages of the class' own API or fields.
  Such usages cause \checkNum{26 (12.2\%)} of the detectors' false positives in their \checkNum{top-20 findings}.
\end{obs}

\vspace{0.03in}
\noindent {\bf 5. Dependent.}
% description
When two objects' states depend upon each other, usages sometimes check the state of one and implicitly draw conclusions about the state of the other.
The detectors do not consider such inter-dependencies.
% example
For example, when two collections are maintained in parallel, i.e., always have the same size, it is sufficient to check the size of one of them before accessing either.
The detectors falsely report \checkNum{14 missing size checks} in such usages.
In \checkNum{10 of these cases}, the equal size is ensured by construction of the collections in the same method.
In the \checkNum{remaining four cases}, it is ensured elsewhere in the same class.
We consider this a dangerous practice, because should the dependency between the collections ever change, it is easy to miss some of the code that relies on it.
Thus, warning developers might be justified.
Nevertheless, we count these cases as false positives, since the current usages are correct.

\begin{obs}{e2-dependent}
  Semantic dependencies between objects' states may implicitly ensure conditions.
  Not considering such inter-dependencies causes \checkNum{14 (6.6\%)} of the detectors' false positives in their \checkNum{top-20 findings}.
\end{obs}

\begin{table*}[tb]
  \centering
  \small
	\caption{\nameref{e1}: Recall of the Isolated Detection Strategies and Root Causes for Divergences.}
    \input{sections/table-ex1-results}
  \label{tab:ex1-results}
\end{table*}

% \paragraph*{Precondition}
% description
% The detectors treat the cooccurrence of facts as a mutual implication, while dependencies between facts may be directed.~\hn{now reading this description, I think this should be in inter-procedural analysis. I would imagine that the method containing this hasNext() will be followed by method containing next() or some other method calls. There is no point calling hasNext() only.}~\sa{In the case of Iterator I agree, however, I think this is a more general pattern, as, for example, it makes sense in some cases to only check the size of a list without iterating or accessing it.}
% example
% For example, all detectors associate \code{hasNext()} and \code{next()}. Consequently, they report actual misuses where only \code{next()} is called. However, they also report false positives where only \code{hasNext()} is called, e.g., to change output depending on wether the iterator is empty or not, without actually consuming it.

\vspace{0.03in}
\noindent {\bf 6. Multiplicity.}
% description
The detectors cannot handle methods that may be called arbitrarily often.
% example
\GROUMiner and \Jadet both learn a pattern where the \code{append()} method of \code{StringBuilder} is called twice and falsely report \checkNum{three missing method calls} where it is called only once.

\begin{obs}{e2-multiplicity}
  Detectors should distinguish methods that require a specific number of calls, from methods that require one or more calls, and methods that may be called arbitrarily often.
  Not considering this causes \checkNum{3 (1.4\%)} of the detectors' false positives in their \checkNum{top-20 findings}.
\end{obs}

\vspace{0.03in}
\noindent {\bf 7. Bug.}
% description
A few findings are likely caused by mistakes in the detector implementations.
% example
\DMMC reports \checkNum{four violations} with an empty set of missing methods.
These empty sets are produced when none of the potentially missing methods match \DMMC's prevalence criteria.
\DMMC should probably filter such empty-set findings before reporting.
\GROUMiner reports \checkNum{one missing \code{if}} that actually appears in all respective usages, because its graph mapping does not match the respective \code{if} node from one of the usages with the corresponding nodes of all the other usages.

\subsection{\nameref{e1}}

We run all detectors to see which of the \checkNum{64 known misuses} from \MUBench they can detect when given the respective crafted correct usages for pattern mining.
\autoref{tab:ex1-results} shows the results per detector.
The second and third columns show the number of potential hits and the number of actual hits, after resolving disagreements.
The fourth and fifth columns show the detectors' empirical recall upper bound and conceptual recall upper bound, respectively.
The sixth column shows the Kappa score for the manual reviews.
The remaining columns show the frequencies of root causes for divergences between a detector's conceptual capabilities from \autoref{tab:detector_capabilities} and its actual findings in this experiment.

We find that \GROUMiner has by far the best recall upper bound and also shows the best recall in \nameref{e1}.
This suggests that its graph representation is a good choice to capture the differences between correct usages and patterns.
However, the gap between \GROUMiner's conceptual upper bound recall and its empirical recall upper bound is quite noticeable.
Actually, \autoref{tab:ex1-results} shows that all four detectors fall considerably short of their conceptual recall upper bound in practice.

Generally, we observe two kinds of divergences between the actual findings and the conceptual capabilities:
\emph{Unexpected false negatives}, i.e., misuses that a detector should be able to~detect, but does not, and \emph{unexpected hits}, i.e., misuses that a detector supposedly cannot detect, but does.
We investigate the root causes of each divergence to identify actionable ways to improve detectors.

\begin{obs}{e1-capability-gap}
  All detectors' empirical recall upper bound is much lower than their conceptual recall upper bound.
  Detectors' findings frequently diverge from their conceptual capabilities.
\end{obs}

The Kappa scores indicate good reviewer agreement, albeit a little lower than in \nameref{e2}.
Since we only reviewed potential hits, i.e., findings in the same method as a known misuse, many potential hits were related to the known misuses.
Consequently, we had several disagreements on whether a particular potential hit actually identifies a particular misuse.
In total, we had \checkNum{18} such disagreements (\Jadet: 4; \GROUMiner: 6; \DMMC: 5; \Tikanga: 3), which led us to formulate the lenient review process described in \autoref{e1}.
We decided in favor of the detectors in \checkNum{eight} of these cases.
We observe that the Kappa score is a little lower for \Jadet, compared to the other detectors.
Since the absolute number of disagreements is comparable and \Jadet had relatively few potential hits, i.e., a small number of decisions as a basis for the Kappa score, we attribute the lower score to chance.
%We decided in favor of \Jadet in \checkNum{three} of the \checkNum{four} cases.

\subsubsection*{Unexpected False Negatives}

%We find the following root causes for unexpected false negatives:

\vspace{0.03in}
\noindent {\bf 1. Representation.}
% description
Current usage representations are not expressive enough to capture all details that are necessary to differentiate between misuses and correct usages.
% example
For example, \DMMC and \GROUMiner encode methods by their name only and, therefore, cannot detect a missing method call, when the usage calls an overloaded version of the respective method.
For example, assume that a pattern requires a call to \code{getBytes(String)}, but the target usage calls \code{getBytes()} instead.
An ideal misuse detector would still report a violation, since the expected method, with the correct parameters, is not called.
However, since only the method name is used for comparison in both these detectors, such a violation is not detected.
Another example is that, to use a \code{Cipher} instance for decryption, it must be in decrypt mode.
This state condition is ensured by passing the constant \code{Cipher.DECRYPT} to the \code{Cipher}'s \code{init()} method.
None of the detectors captures this way of ensuring that the condition holds, because they do not encode method-call arguments in their representations.

\begin{obs}{e1-capture}
 Inability to capture details necessary to differentiate misuses from correct usages in the usage representation is responsible for \checkNum{22 (45.8\%) of the unexpected false negatives}.
\end{obs}

\vspace{0.03in}
\noindent {\bf 2. Matching.}
% description
The detectors fail to relate a pattern and a usage.
Typically, detectors relate patterns and usages by their common facts.
If there are no or only few common facts, detectors report no violation.
% example
For example, \Jadet's facts are pairs of method calls.
In a scenario where \code{JFrame}'s \code{\code{setPreferredSize()}} method is accidentally called after its \code{pack()} method, \Jadet represents the usage with a pair $(\code{pack}, \code{setPreferredSize})$ and the pattern with the reverse pair.
Since it compares facts by equality, \Jadet finds no relation between the pattern and the usage.
Without common facts between a usage and a pattern, the detector assumes that these are two completely unrelated pieces of code and does not report a violation.
Another example is when the pattern's facts relate to a type, e.g., \code{List} in \code{List.size()}, while the usage's facts relate to a super- or sub-type such as \code{ArrayList.size()} or \code{Collection.size()}.
The detectors cannot relate these facts, since they are unaware of the type hierarchy.
Also, \Tikanga misses \checkNum{four misuses}, because the target misses more than two formulae of the pattern (\Tikanga's maximum distance for matching).
For example, \autoref{lst:tikanga-distance} shows a misuse that does not close a \code{Writer} and the corresponding correct usage.
In \Tikanga's representation, the difference between the misuse and the correct usage consists of three formulae:
\begin{enumerate*}[label=(\arabic*)]
  \item that \code{close()} follows \code{write()} in case of normal execution,
  \item that \code{close()} follows \code{write()} if the latter throws an exception, and
  \item that \code{close()} is preceded by a \code{null} check.
\end{enumerate*}

\begin{figure}[tb]
  \begin{subfigure}[t]{0.46\columnwidth}
    \begin{lstlisting}[language=java,numberblanklines=false,escapeinside=||,firstnumber=0]

writer.write(value);|\addtocounter{lstnumber}{-1}|

\end{lstlisting}
  \end{subfigure}
  \begin{subfigure}[t]{0.46\columnwidth}
    \begin{lstlisting}[language=java]
try {
  writer.write(value);
} finally {
  if (writer != null)
    writer.close();
}
    \end{lstlisting}
  \end{subfigure}
  \caption{Not Closing \code{Writer} vs. Correctly Closing \code{Writer}.}
  \label{lst:tikanga-distance}
\end{figure}

\begin{obs}{e1-relate}
  When matching patterns and misuses, detectors should consider the semantics of their representation, e.g., call order and the number of usage facts generated by adding specific usage constructs, as well as code semantics, e.g., subtype relations.
  Neglecting this is responsible for \checkNum{15 (31.3\%) of the unexpected false negatives}.
\end{obs}

\vspace{0.03in}
\noindent {\bf 3. Analysis.}
% description
The detectors rely on static analysis to extract their usage representations.
Imprecisions in these analyses may obscure relations between patterns and usages.
% example
For example, \GROUMiner fails to detect \checkNum{one missing \code{null} check}, because it cannot determine the receiver type for chained calls, such as for \code{m()} in \code{o.getX().m()}, which is not generally possible from source code alone.
Also, it fails to detect another \checkNum{four missing \code{null} checks}, because it overlooks dataflow dependencies.
\autoref{lst:GrouMiner-miss-df} shows such a case.
In addition to the \code{null} check, \GROUMiner also misses the dataflow from the \code{get()} calls to the \code{remove()} call in the misuse, which makes the pattern and usage differ by multiple facts.
\GROUMiner, however, only reports a violation if the difference is a single fact.
\Tikanga misses a call that occurs in the correct usage in \checkNum{one case} and fails to capture the call order between two calls from the correct usage in \checkNum{another case}.
We assume that the cause is a limitation of its analysis, but could not ultimately verify this, because the tool's developer is not available to confirm the implementation details.

\begin{figure}[tb]
  \begin{lstlisting}[language=java]
ArrayList markers;
if (layer == Layer.FOREGROUND) {
  markers = (ArrayList) this.fgMarkers.get(index);
}
else {
  markers = (ArrayList) this.bgMarkers.get(index);
}
// if (markers != null) { // <-- missing in misuse
boolean removed = markers.remove(marker);
// }
  \end{lstlisting}
  \caption{Example of an Analysis Problem of \GROUMiner.}
  \label{lst:GrouMiner-miss-df}
\end{figure}

\begin{obs}{e1-analysis}
  Imprecision of the analysis, which obscures the relation between patterns and misuses, causes \checkNum{9 (18.8\%) of the unexpected false negatives}.
\end{obs}

\vspace{0.03in}
\noindent {\bf 4. Bug.}
% description
%A few false negatives are caused by mistakes in the implementations.
% examples
\DMMC skips the comparison of a usage and a pattern if the pattern contains fewer calls than the usage, presumably to improve performance.
The pattern for \code{AuthState} from Apache's \aName{HTTPClient}, for instance, requires three calls, of which the misuse scenario misses one.
However, if this misuse has an~additional, optional call  that is not in the pattern, \DMMC skips the comparison since now both the pattern and the target each contain 3 calls.
This causes two unexpected false negatives in our experiment.

\subsubsection*{Unexpected Hits}

\vspace{0.03in}
\noindent {\bf 1. Lenient.}
% description
In all but two cases, the reason for \emph{unexpected hits} is the lenient review process we use for \nameref{e1} (see \autoref{e1}).
% example
In most cases, the detectors report a missing call that indicates a missing condition check.
The only other case is that \GROUMiner detects a missing context condition, in a scenario where some \aName{swing} code is required to run on the Event-Dispatching Thread (EDT).
The delegation to the EDT is implemented by wrapping the code in an anonymous instance of \code{Runnable}, as shown in \autoref{lst:edt}.
\GROUMiner considers the code in \code{run()} as part of code of the enclosing method.
Consequently, it suggests the misuse by reporting a missing instantiation of \code{Runnable} before the instantiation of the \code{JFrame}.

\begin{figure}[tb]
  \begin{lstlisting}[language=java]
public static void main(String[] args) {
  SwingUtilities.invokeLater(new Runnable() {
    public void run() {
  	  JFrame f = new JFrame("Main Window");
      // add components...
      f.setVisible(true); 
    }
  });
}
  \end{lstlisting}
  \caption{Instantiating Swing Components on the Event-Dispatching Thread.}
  \label{lst:edt}
\end{figure}

\begin{obs}{e1-lenient}
  Missing method calls may indicate missing condition checks.
  Detectors that report these missing calls, despite not reporting the exact condition, find violations outside of their conceptual capabilities.
\end{obs}

\vspace{0.03in}
\noindent {\bf 2. Exception Handling.}
% description
% example
In the remaining two cases, \Jadet and \Tikanga correctly report missing exception handling.
For example, \autoref{lst:exc-handling} (left) shows a misuse where \code{close()} is not called when \code{write()} throws an exception.
A corresponding correct usage is shown on the right.
\Tikanga and \Jadet both represent the correct usage with two facts $\{ (\code{write}, \code{close}), (\code{write:EXC}, \code{close}) \}$, effectively encoding that \code{close()} is called after \code{write()} in normal execution and in case of an exception.
In the misuse, they find the second fact missing.
This capability of the implementation is not mentioned in the respective publications.

\begin{figure}[tb]
  \begin{subfigure}[t]{0.46\columnwidth}
    \begin{lstlisting}[language=java,numberblanklines=false,escapeinside=||,firstnumber=0]

writer.write(value);|\addtocounter{lstnumber}{-1}|

writer.close();
\end{lstlisting}
  \end{subfigure}
  \begin{subfigure}[t]{0.46\columnwidth}
    \begin{lstlisting}[language=java]
try {
  writer.write(value);
} finally {
  writer.close();
}
    \end{lstlisting}
  \end{subfigure}
  \caption{Closing \code{Writer} Without and With Exception Handling.}
  \label{lst:exc-handling}
\end{figure}

%\begin{obs}{e1-additional}
%  Some detector implementations find missing exception
%  handling, which is beyond the capabilities discussed in the
%  respective publications.
%\end{obs}

%\subsection*{Summary}
%
% However, we identify several unexpected false negatives whose root causes indiciate the need for more expressive usage representations (Capture), code-semantic-aware matching between patterns and targets (Relate), and more precise separation of usages, which entails a precise definition of the boundaries of patterns (Confuse).
%
% On the other hand, we find that missing method calls often indicate missing condition checks.
% Consequently, all detectors unexpectedly indicate misuses outside of their conceptual capabilities (Lenient).
% \sa{I feel we can draw a conclusion here, about cooccurrence of violations, for example, but I cannot put a finger on it right now...}

\subsection{\nameref{e3}}

In \nameref{e3}, we run all detectors to assess their recall without explicitly providing them with correct usages.
In addition to \MUBench's \checkNum{64 misuses}, we add the \checkNum{14 new misuses} from \nameref{e2} and exclude the \checkNum{25 hand-crafted examples} for which there is no project code to mine patterns from.
This leaves us with \checkNum{53 misuses} for \nameref{e3} (Row 5 of~\autoref{tab:datasets}).
%Ideally, a detector would find all these \checkNum{53 misuses}.

\begin{figure}[tb]
  \centering
  \includegraphics[width=.75\columnwidth]{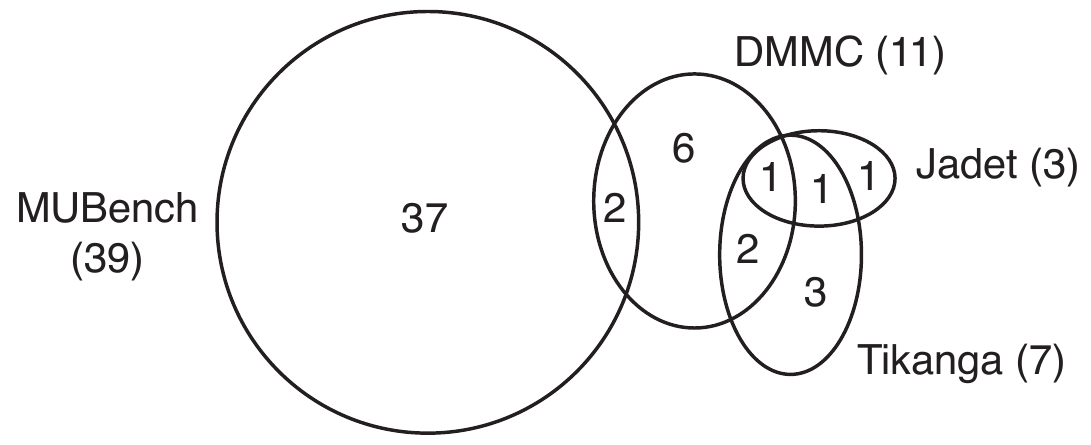}
  \caption{Recall of the Detectors in \nameref{e3}}%~\sn{Really only 2 mubench misuses were found?}~\sa{Yes, sad but true. And those where even discovered by the simplest detector in our experiments...}
  \label{fig:recal-venn}
\end{figure}

\autoref{tab:ex3-results} shows the results and \autoref{fig:recal-venn} visualizes the recall.
\Jadet finds only the \checkNum{three misuses} it already identified in \nameref{e2}.
\GROUMiner does not find any of the misuses.
\Tikanga finds the \checkNum{five misuses} it already identified in \nameref{e2}, \checkNum{one of the misuses} that \DMMC identified in \nameref{e2}, and \checkNum{one of the misuses} that \Jadet identified in \nameref{e2}.
\DMMC finds \checkNum{two misuses} from \MUBench (both missing method calls), the \checkNum{eight misuses} it reported in \nameref{e2}, and \checkNum{one misuse} both \Jadet and \Tikanga reported in \nameref{e2}.

\DMMC shows by far the best recall in \nameref{e3}.
This suggests that its relatively simple detection strategy works well when focusing on missing method calls.
However, the recall of all detectors in the realistic setting offered by \nameref{e3} is low.
Analyzing the root causes for their bad performance, we identify \checkNum{two general problems} with the design of the detectors and their evaluation setup.

\InlineSec{1. Ranking}
\nameref{e3} shows that the detectors identify additional misuses beyond their \checkNum{top-20 findings} that we considered in \nameref{e2}.
Unfortunately, they rank those misuses very low.
For example, the two \MUBench misuses \DMMC finds are ranked \checkNum{309} and \checkNum{613}.
This is far beyond the number of findings that we can reasonably expect a user to assess.
The four detectors in our experiments all use different ranking strategies, but none of the detectors from our survey in \autoref{sec:misuse-detectors-related-work} compared different strategies on the same detector.

\begin{obs}{rank}
  Detectors need better ranking strategies to report true positives within their top findings.
  Furthermore, researchers should compare alternative ranking strategies for single detectors.
\end{obs}

\InlineSec{2. Usage Examples}
 The huge difference in the detectors' performance between Experiments RUB and R suggests that the cause is a shortage of correct usage examples in the target projects.
One possibility is that the number of such examples is smaller than the detectors' minimal support for pattern mining, in which case we could simply lower these thresholds.
However, this would likely also increase the number of false positives as the mined patterns generally become less reliable, which underlines the need to effectively filter false positives (\obsref{e2-precision}) and improve ranking (\obsref{rank}).
Another possibility is that no, or only very few, such examples exist in the projects.
This would be a general problem with the evaluation setup of misuse detectors.
To solve it, we need additional sources of usage examples to mine patterns from.
Gruska~\etal~\cite{GWZ10} demonstrated one possible approach by applying \Jadet in a cross-project setting with 6,000 projects, but did not measure recall.
This strategy is also common in other recommender systems for software engineering, such as code-completion engines~\cite{PLM15}.
The misuse detectors \CARMiner~\cite{TX09} and \Alattin~\cite{TX09b} implement an alternative approach, by specifically searching for usage examples of the APIs used in the target project via a code-search engine.
Related to this, other lines of research proposed code-search engines to find usage examples in open source projects~\cite{GFXM+10,MGPXF11} or on StackOverflow~\cite{PBDO+14}.

\begin{table}[tb]
  \centering
  \small
  \caption{\nameref{e3}: Recall of the Detectors on \MUBench and the New Misuses from \nameref{e2}.}
  \input{sections/table-ex3-results}
  \label{tab:ex3-results}
\end{table}

\begin{obs}{e3-recall}
  All detectors have low recall, likely due to lack of correct usage examples in target projects.
  Adoption of existing code-search techniques and cross-project mining could mitigate this problem.
\end{obs}

The Kappa scores indicate mostly perfect reviewer agreement in \nameref{e3}.
This is because the detectors found almost exclusively the misuses that one of them also identified in \nameref{e2}, i.e., the misuses we already agreed on before.
The exception is \DMMC, where we initially disagreed on \checkNum{one} of its \checkNum{14} potential hits for misuses from the original \MUBench dataset.

%\subsection{Reviewer Agreement}

%Generally, the Kappa scores in all experiments indicate good reviewer agreement across detectors.\footnote{Landis and Koch~\cite{LK77} characterize values from $0.61$ to $0.80$ as substantial and from $0.81$ to $1$ as (almost) perfect agreement.
%Fleiss~\cite{F81} characterizes values over $0.75$ as excellent.}

\subsection{User Experience}

We now report on our experiences as users of our subject misuse detectors.
Our observations is based on the experience we gained while reviewing the detectors' findings in our experiments.

\DMMC simply reports present and missing method calls, along with the source line number of the first present call.
We find this output generally easy to interpret.
The line number helps, especially, to locate usages in large methods.
\GROUMiner reports pattern and usage graphs, which are more difficult to understand.
However, we find that the structural properties of the source code that the graph representation captures help with the interpretation.
\Jadet and \Tikanga report the present and missing facts of their respective representations.
We find that it is often difficult to relate the facts to each other, especially in the presence of multiple usages of the same API.
This might be, in part, due to the textual representation we look at.
While none of the detector implementations was intended to present their findings to end users, we still find it interesting to note that the challenge of explaining findings seems to correlate with the distance between the source code and the usage representation.

% \GROUMiner, which analyzes source code only,  reports a missing \code{toString()} call that is implicit in string concatenation.
% Byte-code based detectors, such as the other three subject detectors in our experiments, do not have these problems, since the compilation emits fully-qualified names and makes all calls explicit.
%
We also find that Bytecode-based detectors may report findings in code that the compiler introduces.
For example, the compiler translates foreach loops into \code{Iterator} usages.
\Tikanga reports a missing call in such a usage, i.e., it reports a missing call on \code{Iterator} in a method where \code{Iterator} does not appear in the source code.
This finding confused us at first.
While additional steps could be taken to assist the user in mapping such findings back to the source code, source-based detectors do not face this problem.

%\begin{obs}{source-vs-bytecode}
%  Source-based detectors may mistakenly report implicit program elements missing, while Bytecode-based detectors may require additional effort to map their findings back to the source~code.
%\end{obs}

Our lenient review process shows that missing method calls frequently indicate missing conditions (\obsref{e1-lenient} and \obsref{e2-misuses-allCSV}).
While such findings do not report the entire problem, we found it relatively easy to deduce their meaning.
In contrast, \GROUMiner reports only a missing \code{if} node, when it captures a missing condition.
While these findings more explicitly indicate the problem of a missing check, we feel that they are actually harder to act upon, because they give no information about \emph{what} should be checked.
This indicates a gap between a detector's capability to find a violation type and its ability to explain respective violations to users.

Above all, we believe that the detectors' precision is likely to be the biggest threat to their applicability in practice.
As a previous study by Johnson~\etal~\cite{JS13} shows, large numbers of false positives are a major barrier in the adoption of code analysis tools.
This problem is made worse by the low recall of the detectors. 
Even if developers would take the time to review all reported warnings, they would still likely miss the vast majority of misuses.

%\begin{obs}{explain}
%  Detectors need mechanisms to explain the misuses to users.
%\end{obs}

\subsection{Call to Action}

We find that misuse detectors are practically capable of detecting a considerable part of the misuses in \MUBench, when provided with the correct usages to compare to (\nameref{e1}).
However, even though the detectors are also capable of finding some misuses in a realistic setting (Experiments P and R), they suffer from extremely low precision (\obsref{e2-precision}) and recall (\obsref{e3-recall}).
We identify \checkNum{four root causes} for false negatives, \checkNum{seven root causes} for false positives, and \checkNum{two general problems} with the design of detectors and how they are typically evaluated.
This leads us to several observations on how to advance the state-of-the-art in API-misuse detection.
Therefore, we call researchers to action:

\begin{itemize}
  % definition of usage
  \item We first need a precise definition of API usages, considering usage properties, such as the usage location~(\obsref{e2-inside}) and call multiplicities~(\obsref{e2-multiplicity}).
  % representation of usage
  \item We need a representation of such usages that captures all code details necessary to distinguish correct usages from misuses (\obsref{e1-capture}) and more precise analyses to identify usages in code (\obsref{e1-analysis} and \obsref{e2-analysis}).
  % pattern mining
  \item We need detectors that retrieve sufficiently many usage examples using project-external sources, such as large project sets or code-search engines~(\obsref{e3-recall}).
  % misuse detection
  \item We need detectors that go beyond the naive assumption that a deviation from the most-frequent usage corresponds to a misuse (\obsref{e2-frequent}), but consider program semantics, such as type hierarchies~(\obsref{e1-relate}) and implicit dependencies between objects (\obsref{e2-dependent}).
  We hypothesize that probabilistic models might be a way to tackle this problem.
  \item We need strategies to properly match patterns and usages in the presence of violations~(\obsref{e1-capability-gap} and \obsref{e1-relate}).
  \item We need strategies to properly handle alternative patterns for the same API (\obsref{e2-alternative}).
  %ranking
  \item Finally, we need good ranking strategies, to reduce the cost of reviewing findings~(\obsref{rank}).
  % studies
\end{itemize}

In order to achieve all this, we need repeatable and replicable studies that enable systematic evaluation and analysis of alternative approaches and strategies.
We publish \MUBench and \MUPipe~\cite{mubench}, as a foundation for such work, and call researchers to use and contribute to this infrastructure, to advance the state of the art in API-misuse detection.

%% file: sections/table-ex2-results.tex
%!TEX root = ../paper.tex

%\setlength{\tabcolsep}{2.5pt}
\small
\begin{tabular}{lrrrrrrrrrrrp{3cm}}
  \toprule
  \multirow{2}{*}{\rot{Detector}} &
  \multirow{2}{*}{\rot{\begin{tabular}[r]{@{}r@{}}Reviewed\\Findings\end{tabular}}} &
  \multirow{2}{*}{\rot{\begin{tabular}[r]{@{}r@{}}Confirmed\\Misuses\end{tabular}}} &
  \multirow{2}{.2in}{\rot{Precision}} &
  \multirow{2}{.2in}{\rot{Kappa Score}} &
  
  \multicolumn{8}{c}{Frequencies of Root Causes for False Positives} \\
  \cmidrule{6-13}
   &  &  &  &  & \rotdia{Uncommon} & \rotdia{Analysis} & \rotdia{Alternative} & \rotdia{Inside} & \rotdia{Dependent} & \rotdia{Bug} & \rotdia{Multiplicity} \\
  \midrule
  \Jadet     &  39 &  4 & 10.3\% & 0.97 &   21 &  3 &  8 &  0 &  1 & 0 & 2 \\
  \GROUMiner &  66 &  0 &  0.0\% & 0.97 &   25 & 22 &  8 &  7 &  2 & 1 & 1 \\
  \DMMC      &  81 &  8 &  9.9\% & 0.91 &    9 & 19 & 18 & 19 &  4 & 4 & 0 \\
  \Tikanga   &  44 &  5 & 11.4\% & 0.93 &   18 &  7 &  7 &  0 &  7 & 0 & 0 \\
  \midrule
  Total      & 230 & 17 &        & 0.94 &   73 & 51 & 41 & 26 & 14 & 5 & 3 \\
  \bottomrule
\end{tabular}

%% file: sections/table-ex1-results.tex
%!TEX root = ../paper.tex

%\setlength{\tabcolsep}{2.5pt}
\small
\begin{tabular}{lrrrrrrrrrrrp{2em}}
  \toprule
  \multirow{2}{*}{\rot{Detector}} &
  \multirow{2}{*}{\rot{Potential Hits}} &
  \multirow{2}{*}{\rot{Actual Hits}} &
  \multirow{2}{*}{\rot{\begin{tabular}[r]{@{}r@{}}Empirical Recall\\ Upper Bound\end{tabular}}} &
  \multirow{2}{*}{\rot{\begin{tabular}[r]{@{}r@{}}Conceptual Recall\\ Upper Bound\end{tabular}}} &
  \multirow{2}{*}{\rot{Kappa Score}} &
  
  \multicolumn{7}{c}{Frequencies of Root Causes} \\
  \cmidrule{7-13}\\[-0.5em] % low extra row to make rotated multirow captions fit
   &  &  &  &  &  & \rotdia{Representation} & \rotdia{Matching} & \rotdia{Analysis} & \rotdia{Bug} & \rotdia{Lenient} & \rotdia{\begin{tabular}[c]{@{}c@{}}Exception\\Handling\end{tabular}} & \\
  \midrule
  \Jadet     & 19 & 15 & 23.4\% & 29.7\% & 0.76 &  4 &  4 & 1 & 0 &  3 & 2 \\
  \GROUMiner & 46 & 31 & 48.4\% & 75.0\% & 0.84 &  9 &  4 & 6 & 0 &  8 & 0 \\
  \DMMC      & 40 & 15 & 23.4\% & 28.1\% & 0.85 &  5 &  0 & 0 & 2 &  5 & 0 \\
  \Tikanga   & 23 & 13 & 20.3\% & 29.7\% & 0.84 &  4 &  7 & 2 & 0 &  5 & 2 \\
  \midrule                                                                
  Total      &    &    &        &        & 0.83 & 22 & 15 & 9 & 2 & 21 & 4 \\
  \bottomrule
  % number of misuses: 64
\end{tabular}

%% file: sections/table-ex3-results.tex
%!TEX root = ../paper.tex

\small
\begin{tabular}{lrrrrrrp{1.2em}}
  \toprule
  Detector & \rotdia{Potential Hits} & \rotdia{Actual Hits} & \rotdia{Recall} & \rotdia{Kappa Score} & \\
  \midrule
  \Jadet     &  4 &  3 &  5.7\% & 1.00 \\
  \GROUMiner &  4 &  0 &  0.0\% & 1.00 \\
  \DMMC      & 25 & 11 & 20.8\% & 0.95 \\
  \Tikanga   &  9 &  7 & 13.2\% & 1.00 \\
  \midrule
  Total      &    &    &        & 0.97 \\
  \bottomrule
\end{tabular}

%% file: sections/threats.tex
%!TEX root = ../paper.tex

\section{Threats to Validity}
\label{sec:threats_to_validity}

%In this section, we discuss possible threats to the validity of our methodology and experiments. 

\InlineSec{Construct Validity}
Any detector's performance is dependent on its configuration. 
%it is possible that the detectors we compared against can perform better on our benchmark, given a different configuration. 
Due to the high effort of reviewing findings, we could not try different configurations for each detector.
However, to give each detector a fair chance, we used the optimal configurations reported in the respective publications.

Our study focuses on static misuse detectors.
Approaches based on dynamic analyses may perform differently and have unique strengths and weaknesses.
To enable dynamic analyses of the project versions in \MUBench, we would have to ensure that the respective code is executable (which requires a sufficient run-time environment, in addition to compile-time dependencies) and to provide example inputs for the execution.
It is unclear how to do this such that it results in a fair comparison of both static and dynamic techniques, without resorting to comparing apples to oranges.
In this work, we focused only on static approaches.

Our experiments focus on detectors that detect misuses in Java code.
Therefore, the results may not generalize to detectors for other languages.
We decided to focus on this subset of detectors, because the majority of approaches we identified in our survey targets Java.
To include detectors that target other languages, we would have to either migrate them to Java or build up additional datasets for the respective languages, both of which is outside the scope of this work.

\InlineSec{Internal Validity}
Reviewing the detectors' findings was done by three of the authors and was not blind (i.e., we knew the detectors we were reviewing findings for).
We could not do blind reviewing, because each approach has a distinct representation of usages and violations that cannot be anonymized.
Moreover, two of the authors of this work are among the original authors of \GROUMiner.
We did our best to review objectively.
To avoid bias, every finding was independently reviewed by two authors and for all findings of \GROUMiner, at least one review was done by an author who was not involved in the original work.

% \sa{Can we roughly quantify our review effort?}~\sn{perhaps using an average of 2min per review is fair? a lot of them took less but some took a bit and if we include discussion then 2min is very fair}~\sa{Taking the number of potential hits ($122 + 230 + 42$) times 2 reviews times 2min/review means $26.3h$ of review work in total. The efforts per detector are 4h for \Jadet, 7.8h for \GROUMiner, 5h for \Tikanga, and 9h for \DMMC. Should we put this as an argument for not involving the original authors?}~\sn{yeah i guess we can. I had a feeling that it took more time than this, but maybe I'm mistaken :D}~\sa{It took more time, because we rereviewed (parts of) the results several times after doing changes, but this is not workload somebody else would have to assess our final results. We also reviewed MuDetect, which is not in here anymore.}
%
While we did ask the original authors to confirm our assessment of the conceptual capabilities of their tools, we did not ask them to confirm the empirical results of our experiments.
We estimate that, including discussions to resolve disagreements, it required each reviewer on average \checkNum{2 minutes to verify whether a detector identified one of the known misuses in Experiments RUB and R} and \checkNum{5 minutes to verify whether a detector's finding identifies an actual misuse in \nameref{e2}}, where we needed to understand the respective code, check documentation, and sometimes also look into transitively called methods.
This amounts to \checkNum{24.8 hours of review effort} per reviewer, \checkNum{4 hours for \Jadet}, \checkNum{7.2 hours for \GROUMiner}, \checkNum{4.7 hours for \Tikanga}, and \checkNum{8.9 hours for \DMMC}.
We decided it is unreasonable to expect the original authors to invest this amount of time in verifying our assessments.
We do, however, publish all our review data~\cite{artifact-page} to allow them and others to revisit our decisions.

% For \Jadet, we measure a similar precision as in the original evaluation (10% vs 9-12%)
% For \Tikanga, we measure a better precision than in the original evaluation (11% vs 6%)
% For \DMMC, we measure a much worse precision than in the original evaluation (9% vs. 85%)

\InlineSec{External Validity}
There may be violation categories we miss in \MUC.
The \MUBench dataset may also not have enough examples of all violations.
This may impact the detectors' comparisons.
However, the existing \MUBench dataset is based on over 1,200 reports from state-of-the-art bug datasets as well as developer input~\cite{ANNN+16} and the results of two empirical studies on API usage directives.
Our survey of existing detectors' capabilities also includes \checkNum{12 detectors}.
This makes it unlikely that we miss a prevalent violation category.

Our dataset may not be representative of all possible real-world API misuses, especially, because we could only compile \checkNum{29 (52\%)} of the \checkNum{55 project versions} and had to exclude the misuses in the other versions from our experiments.
Compiling arbitrary versions of projects from the source control history of project is a challenging task.
We invested two full weeks work of one of the authors and additional 3 months work of a student, to include as many project versions as possible.
Still, loosing the examples for which we could not compile the respective project versions may bias the results of our experiments.

Ideally, our experiments would include thousands of misuses from a large number of projects and in each individual project version, to give us greater confidence in the generalizability of our results.
However, currently, there is no such dataset.
We invested several months of effort to collect and prepare \MUBench in its current state, to make a first step towards a large benchmark.
Now that the we have the infrastructure in place, it is straightforward to extend \MUBench with misuse examples from different sources.

We publish \MUPipe and \MUBench~\cite{mubench} and encourage others to extend the dataset and repeat our experiments, also with other detectors and detector configurations.

%% file: sections/conclusions.tex
%!TEX root = ../paper.tex

\section{Conclusions} % (fold)
\label{sec:conclusions}

API-misuse detectors help developers write better software by warning them about potential misuses in their code.
Despite the existence of many such detectors, there has been no attempt to systematically study types of API misuses and design detectors accordingly.
In this paper, we addressed this gap by creating \MUC, based on a dataset of \checkNum{100 misuses}.
By evaluating the conceptual capabilities of \checkNum{12 existing detectors} against \MUC, we identified shortcomings qualitatively.
We then developed an automated benchmark pipeline, \MUPipe, to empirically evaluate \checkNum{four existing detectors}.
Our results reveal that misuse detectors are practically capable of detecting misuses, when explicitly provided with correct usages to mine patterns from.
However, they suffer from extremely low precision and recall in a realistic application setting.
We identify \checkNum{four root causes} for false negatives, \checkNum{seven root causes} for false positives, and \checkNum{two general problems} with the design of detectors and the commonly-used evaluation setup.
These lead us to several observations on how to advance the state-of-the-art in API-misuse detection in future work.
We publish all our tooling and our dataset~\cite{mubench} to encourage other researchers to join us along this path.

%% file: sections/acknowledgements.tex
%!TEX root = ../paper.tex

\section*{Acknowledgements}

We thank
our students M. Kämmerer and J. Schlitzer for their work on \MUPipe and their help preparing \MUBench,
M. Monperrus for providing \DMMC,
A. Zeller and A. Wasylkowski for providing \Jadet and \Tikanga, and
M. Pradel for additional examples for \MUBench.

This work was partially funded by the German Federal Ministry of Education and Research (BMBF) within the Software Campus project \emph{Eko}, grant no. 01IS12054, by the DFG as part of CRC 1119 CROSSING, and by the Hessen State Ministry for Higher Education, Research and the Arts (HMWK) within CRISP. The authors assume responsibility for the paper content.

%% file: paper.bbl
% Generated by IEEEtran.bst, version: 1.14 (2015/08/26)
\begin{thebibliography}{10}
\providecommand{\url}[1]{#1}
\csname url@samestyle\endcsname
\providecommand{\newblock}{\relax}
\providecommand{\bibinfo}[2]{#2}
\providecommand{\BIBentrySTDinterwordspacing}{\spaceskip=0pt\relax}
\providecommand{\BIBentryALTinterwordstretchfactor}{4}
\providecommand{\BIBentryALTinterwordspacing}{\spaceskip=\fontdimen2\font plus
\BIBentryALTinterwordstretchfactor\fontdimen3\font minus
  \fontdimen4\font\relax}
\providecommand{\BIBforeignlanguage}[2]{{%
\expandafter\ifx\csname l@#1\endcsname\relax
\typeout{** WARNING: IEEEtran.bst: No hyphenation pattern has been}%
\typeout{** loaded for the language `#1'. Using the pattern for}%
\typeout{** the default language instead.}%
\else
\language=\csname l@#1\endcsname
\fi
#2}}
\providecommand{\BIBdecl}{\relax}
\BIBdecl

\bibitem{MM13}
M.~Monperrus and M.~Mezini, ``Detecting missing method calls as violations of
  the majority rule,'' \emph{{ACM} Transactions on Software Engineering and
  Methodology}, vol.~22, no.~1, pp. 1--25, 2013.

\bibitem{SHA15}
J.~Sushine, J.~D. Herbsleb, and J.~Aldrich, ``Searching the state space: {A}
  qualitative study of {API} protocol usability,'' in \emph{Proceedings of the
  23\textsuperscript{rd} {IEEE} International Conference on Program
  Comprehension}, ser. ICPC '15.\hskip 1em plus 0.5em minus 0.4em\relax {IEEE}
  Computer Society Press, 2015, pp. 82--93.

\bibitem{ANNN+16}
S.~Amann, S.~Nadi, H.~A. Nguyen, T.~N. Nguyen, and M.~Mezini, ``{MUBench}: {A}
  benchmark for {API}-misuse detectors,'' in \emph{Proceedings of the
  13\textsuperscript{th} Working Conference on Mining Software Repositories},
  ser. MSR '16.\hskip 1em plus 0.5em minus 0.4em\relax {ACM} Press, 2016.

\bibitem{FHMB+12}
S.~Fahl, M.~Harbach, T.~Muders, L.~Baumg\"{a}rtner, B.~Freisleben, and
  M.~Smith, ``Why {Eve} and {Mallory} love {Android}: {A}n analysis of {Android
  SSL} (in)security,'' in \emph{Proceedings of the 19\textsuperscript{th} {ACM}
  Conference on Computer and Communications Security}, ser. CCS '12.\hskip 1em
  plus 0.5em minus 0.4em\relax {ACM} Press, 2012, pp. 50--61.

\bibitem{EBFK13}
M.~Egele, D.~Brumley, Y.~Fratantonio, and C.~Kruegel, ``An empirical study of
  cryptographic misuse in {Android} applications,'' in \emph{Proceedings of the
  Conference on Computer \& Communications Security}, ser. CCS'13.\hskip 1em
  plus 0.5em minus 0.4em\relax {ACM} Press, 2013, pp. 73--84.

\bibitem{NKMB16}
S.~Nadi, S.~Kr\"{u}ger, M.~Mezini, and E.~Bodden, ``"{Jumping} through hoops":
  {W}hy do developers struggle with cryptography {APIs}?'' in \emph{Proceedings
  of the 38\textsuperscript{th} International Conference on Software
  Engineering}, ser. ICSE'16.\hskip 1em plus 0.5em minus 0.4em\relax {ACM}
  Press, 2016.

\bibitem{GIJA+12}
M.~Georgiev, S.~Iyengar, S.~Jana, R.~Anubhai, D.~Boneh, and V.~Shmatikov, ``The
  most dangerous code in the world: {V}alidating {SSL} certificates in
  non-browser software,'' in \emph{Proceedings of the 19\textsuperscript{th}
  {ACM} Conference on Computer and Communications Security}, ser. CCS
  '12.\hskip 1em plus 0.5em minus 0.4em\relax {ACM} Press, 2012, pp. 38--49.

\bibitem{DH09}
U.~Dekel and J.~D. Herbsleb, ``Improving {API} documentation usability with
  knowledge pushing,'' in \emph{Proceedings of the 31\textsuperscript{st}
  International Conference on Software Engineering}, ser. ICSE '09.\hskip 1em
  plus 0.5em minus 0.4em\relax {IEEE} Computer Society Press, 2009, pp.
  320--330.

\bibitem{ABFKMS16}
Y.~Acar, M.~Backes, S.~Fahl, D.~Kim, M.~L. Mazurek, and C.~Stransky, ``You get
  where you're looking for. {T}he impact of information sources on code
  security,'' in \emph{Proceedings of the 37\textsuperscript{th} {IEEE}
  Symposium on Security and Privacy}.\hskip 1em plus 0.5em minus 0.4em\relax
  {IEEE} Computer Society Press, 2016.

\bibitem{artifact-page}
\BIBentryALTinterwordspacing
``{Artifact Page},'' 2017. [Online]. Available:
  \url{http://www.st.informatik.tu-darmstadt.de/artifacts/mustudy/}
\BIBentrySTDinterwordspacing

\bibitem{LZ05}
Z.~Li and Y.~Zhou, ``{PR-Miner}: {A}utomatically extracting implicit
  programming rules and detecting violations in large software code,'' in
  \emph{Proceedings of the 10\textsuperscript{th} European Software Engineering
  Conference Held Jointly with 13\textsuperscript{th} ACM SIGSOFT International
  Symposium on Foundations of Software Engineering}, ser. ESEC/FSE '13.\hskip
  1em plus 0.5em minus 0.4em\relax {ACM} Press, 2005, pp. 306--315.

\bibitem{L07}
C.~Lindig, ``Mining patterns and violations using concept analysis,''
  Universit{\"a}t des Saarlandes, Saarbr{\"u}cken, Germany, Tech. Rep., 2007.

\bibitem{WZL07}
A.~Wasylkowski, A.~Zeller, and C.~Lindig, ``Detecting object usage anomalies,''
  in \emph{Proceedings of the 6\textsuperscript{th} {ACM} Joint Meeting of the
  European Software Engineering Conference and the {ACM SIGSOFT} Symposium on
  The Foundations of Software Engineering}, ser. ESEC/FSE '07.\hskip 1em plus
  0.5em minus 0.4em\relax {ACM} Press, 2007, pp. 35--44.

\bibitem{RGJ07}
M.~K. Ramanathan, A.~Grama, and S.~Jagannathan, ``Static specification
  inference using predicate mining,'' in \emph{Proceedings of the
  28\textsuperscript{th} {ACM SIGPLAN} Conference on Programming Language
  Design and Implementation}, ser. PLDI '07.\hskip 1em plus 0.5em minus
  0.4em\relax {ACM} Press, 2007, pp. 123--134.

\bibitem{NNP+09}
T.~T. Nguyen, H.~A. Nguyen, N.~H. Pham, J.~M. Al-Kofahi, and T.~N. Nguyen,
  ``Graph-based mining of multiple object usage patterns,'' in
  \emph{Proceedings of the 7\textsuperscript{th} {ACM} Joint Meeting of the
  European Software Engineering Conference and the {ACM SIGSOFT} Symposium on
  The Foundations of Software Engineering}, ser. ESEC/FSE '09.\hskip 1em plus
  0.5em minus 0.4em\relax {ACM} Press, 2009, pp. 383--392.

\bibitem{AX09}
M.~Acharya and T.~Xie, ``Mining {API} error-handling specifications from source
  code,'' in \emph{Proceedings of the 12\textsuperscript{th} International
  Conference on Fundamental Approaches to Software Engineering: Held As Part of
  the Joint European Conferences on Theory and Practice of Software, ETAPS
  2009}, ser. FASE '09.\hskip 1em plus 0.5em minus 0.4em\relax {Springer-Verlag
  GmbH}, 2009, pp. 370--384.

\bibitem{TX09}
S.~Thummalapenta and T.~Xie, ``Mining exception-handling rules as sequence
  association rules,'' in \emph{Proceedings of the 31\textsuperscript{st}
  International Conference on Software Engineering}, ser. ICSE '09.\hskip 1em
  plus 0.5em minus 0.4em\relax {IEEE} Computer Society Press, 2009, pp.
  496--506.

\bibitem{TX09b}
------, ``{Alattin}: {M}ining alternative patterns for detecting neglected
  conditions,'' in \emph{Proceedings of the 24\textsuperscript{th} {IEEE/ACM}
  International Conference on Automated Software Engineering}, ser. ASE
  '09.\hskip 1em plus 0.5em minus 0.4em\relax {IEEE} Computer Society Press,
  2009, pp. 283--294.

\bibitem{WZ11}
A.~Wasylkowski and A.~Zeller, ``Mining temporal specifications from object
  usage,'' \emph{Automated Software Engineering}, vol.~18, no. 3-4, pp.
  263--292, 2011.

\bibitem{NPVN16}
T.~T. Nguyen, H.~V. Pham, P.~M. Vu, and T.~T. Nguyen, ``Recommending {API}
  usages for mobile apps with {Hidden Markov Model},'' in \emph{Proceedings of
  the 30\textsuperscript{th} {ACM/IEEE} International Conference on Automated
  Software Engineering}, ser. ASE '15.\hskip 1em plus 0.5em minus 0.4em\relax
  {IEEE} Computer Society Press, 2015, pp. 795--800.

\bibitem{LHXRM16}
O.~Legunsen, W.~U. Hassan, X.~Xu, G.~Ro\c{s}u, and D.~Marinov, ``How good are
  the specs? {A} study of the bug-finding effectiveness of existing {Java}
  {API} specifications,'' in \emph{Proceedings of the 31\textsuperscript{st}
  {IEEE/ACM} International Conference on Automated Software Engineering}, ser.
  ASE '16.\hskip 1em plus 0.5em minus 0.4em\relax {ACM} Press, 2016, pp.
  602--613.

\bibitem{METM12}
M.~Monperrus, M.~Eichberg, E.~Tekes, and M.~Mezini, ``What should developers be
  aware of? {A}n empirical study on the directives of {API} documentation,''
  \emph{Empirical Software Engineering}, vol.~17, no.~6, pp. 703--737, 2012.

\bibitem{farewellGoogle}
\BIBentryALTinterwordspacing
C.~DiBona, ``Bidding farewell to {Google Code}.'' [Online]. Available:
  \url{http://google-opensource.blogspot.com/2015/03/farewell-to-google-code.html}
\BIBentrySTDinterwordspacing

\bibitem{mubench}
\BIBentryALTinterwordspacing
``{MUBench},'' 2017. [Online]. Available:
  \url{https://github.com/stg-tud/MUBench/}
\BIBentrySTDinterwordspacing

\bibitem{PG12}
M.~Pradel and T.~R. Gross, ``Leveraging test generation and specification
  mining for automated bug detection without false positives,'' in
  \emph{Proceedings of the 34\textsuperscript{th} International Conference on
  Software Engineering}, ser. ICSE '12.\hskip 1em plus 0.5em minus 0.4em\relax
  {IEEE} Computer Society Press, 2012, pp. 288--298.

\bibitem{LZLJMSR14}
Q.~Luo, Y.~Zhang, C.~Lee, D.~Jin, P.~O. Meredith, T.~F.
  {\c{S}}erb{\u{a}}nu{\c{t}}{\u{a}}, and G.~Rosu, ``{RV-Monitor}: {E}fficient
  parametric runtime verification with simultaneous properties,'' in
  \emph{Runtime Verification}.\hskip 1em plus 0.5em minus 0.4em\relax
  {Springer-Verlag GmbH}, 2014, pp. 285--300.

\bibitem{PJAG12}
M.~Pradel, C.~Jaspan, J.~Aldrich, and T.~R. Gross, ``Statically checking {API}
  protocol conformance with mined multi-object specifications,'' in
  \emph{Proceedings of the 34\textsuperscript{th} International Conference on
  Software Engineering}, ser. ICSE '12.\hskip 1em plus 0.5em minus 0.4em\relax
  {IEEE} Computer Society Press, 2012, pp. 925--935.

\bibitem{ECHC+01}
D.~Engler, D.~Y. Chen, S.~Hallem, A.~Chou, and B.~Chelf, ``Bugs as deviant
  behavior: {A} general approach to inferring errors in systems code,'' in
  \emph{Proceedings of the 18\textsuperscript{th} {ACM} Symposium on Operating
  Systems Principles}, ser. SOSP '01.\hskip 1em plus 0.5em minus 0.4em\relax
  {ACM} Press, 2001, pp. 57--72.

\bibitem{IEEE10}
``{IEEE} standard classification for software anomalies,'' \emph{IEEE Std
  1044-2009 (Revision of IEEE Std 1044-1993)}, pp. 1--23, 2010.

\bibitem{CBCHMRW92}
R.~Chillarege, I.~S. Bhandari, J.~K. Chaar, M.~J. Halliday, D.~S. Moebus, B.~K.
  Ray, and M.-Y. Wong, ``Orthogonal defect classification-a concept for
  in-process measurements,'' \emph{{IEEE} Transactions on Software
  Engineering}, vol.~18, no.~11, pp. 943--956, 1992.

\bibitem{BBMZ16}
M.~Beller, R.~Bholanath, S.~McIntosh, and A.~Zaidman, ``Analyzing the state of
  static analysis: {A} large-scale evaluation in open source software,'' in
  \emph{Proceedings of the 23\textsuperscript{rd} {IEEE} International
  Conference on Software Analysis, Evolution, and Reengineering}, vol.~1.\hskip
  1em plus 0.5em minus 0.4em\relax {IEEE} Computer Society Press, 2016.

\bibitem{GS67}
B.~G. Glaser and A.~L. Strauss, \emph{The Discovery of Grounded Theory:
  {S}trategies for Qualitative Research}.\hskip 1em plus 0.5em minus
  0.4em\relax Aldine de Gruyter, 1967, vol.~46.

\bibitem{RBMR13}
M.~P. Robillard, E.~Bodden, D.~Kawrykow, M.~Mezini, and T.~Ratchford,
  ``Automated {API} property inference techniques,'' \emph{{IEEE} Transactions
  on Software Engineering}, vol.~39, no.~5, pp. 613--637, 2013.

\bibitem{RGJ07b}
M.~K. Ramanathan, A.~Grama, and S.~Jagannathan, ``Path-sensitive inference of
  function precedence protocols,'' in \emph{Proceedings of the
  29\textsuperscript{th} International Conference on Software Engineering},
  ser. ICSE '07.\hskip 1em plus 0.5em minus 0.4em\relax {IEEE} Computer Society
  Press, 2007, pp. 240--250.

\bibitem{GW99}
B.~Ganter and R.~Wille, \emph{Formal Concept Analysis: {M}athematical
  Foundations}, 1st~ed.\hskip 1em plus 0.5em minus 0.4em\relax {Springer-Verlag
  New York, Inc.}, 1997.

\bibitem{MBM10}
M.~Monperrus, M.~Bruch, and M.~Mezini, ``Detecting missing method calls in
  object-oriented software,'' in \emph{Proceedings of the
  24\textsuperscript{th} European Conference on Object-oriented Programming},
  ser. ECOOP '10.\hskip 1em plus 0.5em minus 0.4em\relax {Springer-Verlag
  GmbH}, 2010, pp. 2--25.

\bibitem{GWZ10}
N.~Gruska, A.~Wasylkowski, and A.~Zeller, ``Learning from 6,000 projects,'' in
  \emph{Proceedings of the 19\textsuperscript{th} International Symposium on
  Software Testing and Analysis}, ser. ISTA '10.\hskip 1em plus 0.5em minus
  0.4em\relax {ACM} Press, 2010, pp. 119--129.

\bibitem{BMUT97}
S.~Brin, R.~Motwani, J.~D. Ullman, and S.~Tsur, ``Dynamic itemset counting and
  implication rules for market basket data,'' in \emph{Proceedings of the {ACM
  SIGMOD} International Conference on Management of Data}, ser. SIGMOD
  '97.\hskip 1em plus 0.5em minus 0.4em\relax {ACM} Press, 1997, pp. 255--264.

\bibitem{FLLNSSLNSS02}
C.~Flanagan, K.~R.~M. Leino, M.~Lillibridge, G.~Nelson, J.~B. Saxe, R.~Stata,
  M.~Lillibridge, G.~Nelson, J.~B. Saxe, and R.~Stata, ``Extended static
  checking for {Java},'' \emph{{ACM SIGPLAN} Notices}, vol.~37, no.~5, pp.
  234--245, 2002.

\bibitem{BBCCFHHKME10}
A.~Bessey, K.~Block, B.~Chelf, A.~Chou, B.~Fulton, S.~Hallem, C.~Henri-Gros,
  A.~Kamsky, S.~McPeak, and D.~Engler, ``A few billion lines of code later:
  {U}sing static analysis to find bugs in the real world,''
  \emph{Communications of the {ACM}}, vol.~53, no.~2, pp. 66--75, 2010.

\bibitem{JS13}
B.~Johnson and Y.~Song, ``Why don't software developers use static analysis
  tools to find bugs?'' in \emph{Proceedings of the 35\textsuperscript{th}
  International Conference on Software Engineering}, ser. ICSE '13.\hskip 1em
  plus 0.5em minus 0.4em\relax {IEEE} Computer Society Press, 2013.

\bibitem{closure}
\BIBentryALTinterwordspacing
G.~Inc., ``Closure compiler,'' 2017. [Online]. Available:
  \url{https://developers.google.com/closure/compiler/}
\BIBentrySTDinterwordspacing

\bibitem{itext}
\BIBentryALTinterwordspacing
iText Software, ``{iText, a Java PDF Library},'' 2017. [Online]. Available:
  \url{https://sourceforge.net/projects/itext/}
\BIBentrySTDinterwordspacing

\bibitem{jmrtd}
\BIBentryALTinterwordspacing
R.~U. i.~N. Digital Security~group, ``{JMRTD: An Open Source Java
  Implementation of Machine Readable Travel Documents},'' 2017. [Online].
  Available: \url{http://jmrtd.org/}
\BIBentrySTDinterwordspacing

\bibitem{jodatime}
\BIBentryALTinterwordspacing
T.~J. project, ``{Joda-Time},'' 2017. [Online]. Available:
  \url{http://www.joda.org/joda-time/}
\BIBentrySTDinterwordspacing

\bibitem{lucene}
\BIBentryALTinterwordspacing
T.~A.~S. Foundation, ``{Apache Lucene},'' 2017. [Online]. Available:
  \url{https://lucene.apache.org/core/}
\BIBentrySTDinterwordspacing

\bibitem{LLQ+05}
S.~Lu, Z.~Li, F.~Qin, L.~Tan, P.~Zhou, and Y.~Zhou, ``{BugBench}: {B}enchmarks
  for evaluating bug detection tools,'' in \emph{In Workshop on the Evaluation
  of Software Defect Detection Tools}, 2005.

\bibitem{CHKL+09}
C.~Cifuentes, C.~Hoermann, N.~Keynes, L.~Li, S.~Long, E.~Mealy, M.~Mounteney,
  and B.~Scholz, ``{BegBunch}: Benchmarking for {C} bug detection tools,'' in
  \emph{Proceedings of the International Workshop on Defects in Large Software
  Systems}, ser. DEFECTS '09.\hskip 1em plus 0.5em minus 0.4em\relax {ACM}
  Press, 2009, pp. 16--20.

\bibitem{SBA+15}
M.~A. Saied, O.~Benomar, H.~Abdeen, and H.~Sahraoui, ``Mining multi-level {API}
  usage patterns,'' in \emph{Proceedings of the 22\textsuperscript{nd} {IEEE}
  International Conference on Software Analysis, Evolution, and Reengineering},
  ser. SANER '15.\hskip 1em plus 0.5em minus 0.4em\relax {IEEE} Computer
  Society Press, 2015, pp. 23--32.

\bibitem{PLM15}
S.~Proksch, J.~Lerch, and M.~Mezini, ``Intelligent code completion with
  {Bayesian} networks,'' \emph{{ACM} Transactions on Software Engineering and
  Methodology (TOSEM)}, vol.~25, no.~1, pp. 1--31, 2015.

\bibitem{GFXM+10}
M.~Grechanik, C.~Fu, Q.~Xie, C.~McMillan, D.~Poshyvanyk, and C.~Cumby, ``A
  search engine for finding highly relevant applications,'' in
  \emph{Proceedings of the 32\textsuperscript{nd} {ACM/IEEE} International
  Conference on Software Engineering - Volume 1}, ser. ICSE '10.\hskip 1em plus
  0.5em minus 0.4em\relax {ACM} Press, 2010, pp. 475--484.

\bibitem{MGPXF11}
C.~McMillan, M.~Grechanik, D.~Poshyvanyk, Q.~Xie, and C.~Fu, ``Portfolio:
  {F}inding relevant functions and their usage,'' in \emph{Proceedings of the
  33\textsuperscript{rd} International Conference on Software Engineering},
  ser. ICSE '11.\hskip 1em plus 0.5em minus 0.4em\relax {ACM} Press, 2011, pp.
  111--120.

\bibitem{PBDO+14}
L.~Ponzanelli, G.~Bavota, M.~Di~Penta, R.~Oliveto, and M.~Lanza, ``Mining
  {StackOverflow} to turn the {IDE} into a self-confident programming
  prompter,'' in \emph{Proceedings of the 11\textsuperscript{th} Working
  Conference on Mining Software Repositories}, ser. MSR '14.\hskip 1em plus
  0.5em minus 0.4em\relax {ACM} Press, 2014, pp. 102--111.

\end{thebibliography}
